\documentclass[aip,apl,amsmath, twocolumn, amssymb, reprint]{revtex4-2}
\usepackage[T1]{fontenc}
\usepackage{graphicx}
\usepackage{changepage}
\usepackage{xcolor}
\usepackage{dcolumn}
\usepackage{comment}
\usepackage{amsmath,amssymb,bbold,bm}
\usepackage{hyperref}
\usepackage{amsfonts}
\usepackage{dsfont}
\usepackage{listings}
\usepackage{braket}
\usepackage{float,latexsym}
\usepackage{multirow}
\usepackage{tikz}
\usepackage{color}
\usepackage{xr}
\usepackage[normalem]{ulem}
\usepackage{bibunits}
\usepackage{cleveref}
\usepackage{physics}
\usepackage{blindtext}
\usepackage{comment}
\usepackage{tabularx}
\usepackage{ltablex}
\usepackage{array}
\usepackage{pifont}
\usepackage[caption=false]{subfig}
\usepackage{siunitx}
\usepackage{booktabs}
\usepackage{threeparttable}
\usepackage{circuitikz}

\makeatletter

\def\env@sqcases{%
  \let\@ifnextchar\new@ifnextchar
  \left\lbrack
  \def\arraystretch{1.2}%
  \array{@{}l@{\quad}l@{}}%
}
\makeatother

\crefname{appendix}{Appendix}{Appendices}
\crefname{equation}{Eq.}{Eqs.}
\crefname{figure}{Fig.}{Figs.}
\crefname{table}{Table}{Tables}
\crefname{section}{Section}{Sections}
\crefname{enumi}{Case}{Cases}
\creflabelformat{appendix}{[#2#1#3]}
\crefrangelabelformat{figure}{#3#1#4--(#5\crefstripprefix{#1}{#2}#6}
\crefmultiformat{figure}{Figs.~#2#1\xdef\mycreffirstarg{#1}#3}%
 { and~#2#1#3}{, #2#1#3}{ and~#2#1#3}

\def\ie{{\it i.e.}\ }
\def\eg{{\it e.g.}\ }

\newcolumntype{M}[1]{>{\centering\arraybackslash}m{#1}}

\makeatother

\begin{document}
\title{Low-Noise Quantum Dots in Ultra-Shallow Ge/SiGe Heterostructures for Prototyping Hybrid Semiconducting-Superconducting Devices}
\author{Maksim Borovkov}
\email{maksim.borovkov@ist.ac.at}
\affiliation{ISTA,~Institute~of~Science~and~Technology~Austria,~Am~Campus~1,~3400~Klosterneuburg,~Austria}
\author{Yona Schell}
\affiliation{ISTA,~Institute~of~Science~and~Technology~Austria,~Am~Campus~1,~3400~Klosterneuburg,~Austria}
\author{Dina Sokolova}
\affiliation{ISTA,~Institute~of~Science~and~Technology~Austria,~Am~Campus~1,~3400~Klosterneuburg,~Austria}
\author{Kevin Roux}
\affiliation{ISTA,~Institute~of~Science~and~Technology~Austria,~Am~Campus~1,~3400~Klosterneuburg,~Austria}
\author{Paul Falthansl-Scheinecker}
\affiliation{ISTA,~Institute~of~Science~and~Technology~Austria,~Am~Campus~1,~3400~Klosterneuburg,~Austria}
\author{Giorgio Fabris}
\affiliation{ISTA,~Institute~of~Science~and~Technology~Austria,~Am~Campus~1,~3400~Klosterneuburg,~Austria}
\author{Devashish Shah}
\affiliation{ISTA,~Institute~of~Science~and~Technology~Austria,~Am~Campus~1,~3400~Klosterneuburg,~Austria}
\author{Jaime Saez-Mollejo}
\affiliation{ISTA,~Institute~of~Science~and~Technology~Austria,~Am~Campus~1,~3400~Klosterneuburg,~Austria}
\author{Rodolfo Previdi}
\affiliation{ISTA,~Institute~of~Science~and~Technology~Austria,~Am~Campus~1,~3400~Klosterneuburg,~Austria}
\author{Inas Taha}
\affiliation{Catalan Institute of Nanoscience and Nanotechnology – ICN2 (CSIC and BIST), Campus UAB, Bellaterra, 08193 Barcelona, Catalonia, Spain
}

\author{Aziz Gen\c{c}}
\affiliation{Catalan Institute of Nanoscience and Nanotechnology – ICN2 (CSIC and BIST), Campus UAB, Bellaterra, 08193 Barcelona, Catalonia, Spain
}
\author{Jordi Arbiol}
\affiliation{Catalan Institute of Nanoscience and Nanotechnology – ICN2 (CSIC and BIST), Campus UAB, Bellaterra, 08193 Barcelona, Catalonia, Spain
}
\affiliation{ICREA, Passeig de Llu´ıs Companys 23, 08010 Barcelona, Catalonia, Spain
}
\author{Stefano Calcaterra}
\affiliation{L-NESS, Physics Department, Politecnico di Milano, via Anzani 42, 22100, Como, Italy}
\author{Afonso De Cerdeira Oliveira }
\affiliation{L-NESS, Physics Department, Politecnico di Milano, via Anzani 42, 22100, Como, Italy}
\author{Daniel Chrastina}
\affiliation{L-NESS, Physics Department, Politecnico di Milano, via Anzani 42, 22100, Como, Italy}
\author{Giovanni Isella}
\affiliation{L-NESS, Physics Department, Politecnico di Milano, via Anzani 42, 22100, Como, Italy}
\author{Anton Bubis}
\affiliation{ISTA,~Institute~of~Science~and~Technology~Austria,~Am~Campus~1,~3400~Klosterneuburg,~Austria}
\author{Georgios Katsaros}
\email{georgios.katsaros@ist.ac.at}
\affiliation{ISTA,~Institute~of~Science~and~Technology~Austria,~Am~Campus~1,~3400~Klosterneuburg,~Austria}

\date{\today}

\begin{abstract}

Planar germanium is currently the only semiconducting platform where high-coherence spin qubits and proximity-induced superconductivity have each been demonstrated. Recent research into spin qubits in Ge/SiGe heterostructures has focused on increasing the thickness of the SiGe capping layer, reporting improvements in the electrostatic noise levels. Meanwhile, heterostructures with thinner capping layers remain rather unexplored, despite the potential advantages for proximity-induced superconductivity. Here, we study a Ge/SiGe heterostructure with a thin SiGe cap $d \approx \SI{4}{\nano \meter}$ and investigate its viability to host low noise quantum dots. To keep the thermal budget compatible with superconducting layers, low-temperature oxide deposition processes were developed and implemented for the gate dielectrics. The charge noise level of fabricated devices is estimated to be $1.8 \pm \SI{1.0}{\micro \eV}/{\sqrt{\SI{}{\Hz}}}$, comparable to devices fabricated on shallow heterostructures ($ d \sim \SI{20}{\nano \meter}$) with high-temperature deposited oxides. Low charge-noise levels, together with the straightforward integration of superconductors, make this heterostructure an attractive platform for prototyping hybrid semiconducting–superconducting devices.
\end{abstract}

\maketitle

Quantum devices based on hybrid semiconductor-superconductor technology have been gaining increasing attention in the past decade, strongly motivated by proposals related to topological superconductivity~\cite{aguado_majorana_2017} and novel types of spin qubits\cite{pita-vidal_novel_2025}. So far, most of the research has focused on group III-V semiconductors, with prominent directions including studies of proximity-induced subgap states \cite{prada_andreev_2020}, Cooper Pair Splitters\cite{wang_singlet_2022, wang_triplet_2023} and engineered Kitaev chains\cite{dvir_realization_2023, ten_haaf_two-site_2024}. However, hybrid devices that use the spin degree of freedom, e.g., Andreev Spin Qubits \cite{hays_coherent_2021, pita-vidal_direct_2023}, exhibit short coherence times due to the interaction with the nuclear spins of the host lattice. This motivates the search for another semiconducting platform with low hyperfine interaction and compatibility with proximity-induced superconductivity. Ge/SiGe heterostructures \cite{scappucci_germanium_2021} are such a platform, as they can host spin qubits with long coherence times \cite{hendrickx_sweet-spot_2024, jirovec_singlet-triplet_2021} and allow for induced superconducting order \cite{hendrickxGatecontrolledQuantumDots2018a, vigneau_germanium_2019, aggarwal_enhancement_2021, tosato_hard_2023, valentini_parity-conserving_2024}. 

Existing Ge/SiGe heterostructures typically place the Ge quantum well
$\sim \SI{20}{\nano\meter}$--$\SI{100}{\nano\meter}$ below the surface\cite{sammak_shallow_2019}. It has been demonstrated that the charge noise levels -- one of the limiting factors for hole spin qubit coherence times\cite{wang_dephasing_2025} -- reduce from the level of $\SI{1.4}{\micro \eV}/{\sqrt{\SI{}{\Hz}}}$ for a shallow heterostructure ($d \approx\SI{22}{\nano\meter}$)\cite{hendrickxGatecontrolledQuantumDots2018a} to $\SI{0.6}{\micro \eV}/{\sqrt{\SI{}{\Hz}}}$ in a deeper $d \approx \SI{55}{\nano\meter}$ heterostructure\cite{lodariLowPercolationDensity2021b}. This finding suggests that thicker capping layers are preferable, provided that quantum dots (QDs) can still be electrostatically defined. For shallow or deep heterostructures, the procedure of bringing a superconductor into the proximity to the quantum well typically relies on either annealing\cite{hendrickxGatecontrolledQuantumDots2018a, tosato_hard_2023, lakic_quantum_2025} or etching\cite{aggarwal_enhancement_2021, vigneau_germanium_2019} through the cap. If, however, the capping layer is sufficiently thin ($\approx \SI{4}{\nano \meter}$), proximity-induced superconductivity can be achieved by directly depositing a superconducting film onto the surface of the heterostructure after native oxide removal\cite{valentini_parity-conserving_2024}. A drawback of this approach is that it brings the quantum well much closer to the surface, potentially increasing electrostatic noise levels due to coupling to interface and oxide traps. As such, the approach raises a key question: can an ultra-shallow ($\approx \SI{4}{\nano \meter}$) Ge/SiGe heterostructure host QDs with charge noise levels suitable for spin qubits with long coherence times?

\begin{figure*}[t]
  \centering
  \includegraphics[width=\textwidth]{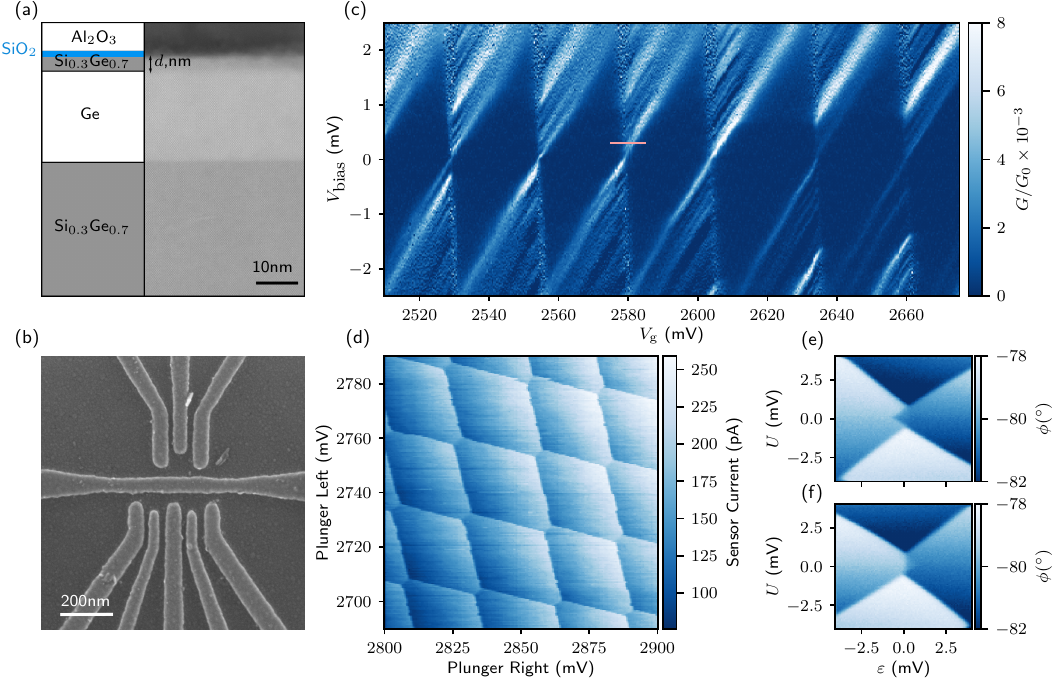}
  \caption{QDs on an ultra-shallow Ge/SiGe heterostructure. (a) HAADF-STEM of the Ge/SiGe heterostructure with the Al$_2$O$_3$ oxide deposited on top. The estimated cap thickness is $d \approx \SI{4}{\nano \meter}$. (b) SEM of a device lithographically identical to Device A. (c) Coulomb diamonds of the left QD. $V_{\mathrm{bias}}$ is the external voltage and $G$ is the conductance calculated as a numerical derivative of the current $I$. The presented data is not corrected for the series resistance $R_{\mathrm{in}}$. Pink cut indicates the gate voltage span over which the time traces in \cref{fig:figure2} were taken. (d) DQD stability diagram measured with the charge sensor. (e-f) Fast scans over the virtual detuning $\varepsilon$ and energy $U$ axes of the DQD. The charge occupation of the dots is measured by using a reflectometry circuit analogous to the one used in Ref\cite{jirovec_singlet-triplet_2021}. An out-of-plane magnetic field $B_{\perp} \approx \SI{5}{\milli \tesla}$ was applied.}
  \label{fig:figure1}
\end{figure*}

Beyond the considerations of the capping layer thickness, a hybrid platform introduces another challenge related to the reduction of the thermal budget once the superconducting layer is deposited. A standard spin qubit fabrication flow involves the atomic layer deposition (ALD) of gate oxide, typically performed at $\approx \SI{300}{\celsius}$\cite{lawrie_quantum_2020, jirovec_singlet-triplet_2021}. Such high temperatures can cause undesired interdiffusion and reactions at the superconductor–semiconductor interface\cite{sistaniMonolithicAxialRadial2018}. To maintain control over the superconducting layer during fabrication, low-temperature ALD ($<\SI{150}{\celsius}$) is therefore required. A few studies have already demonstrated QDs with reduced temperature ALD oxides \cite{lakic_quantum_2025, kulesh_quantum_2020}. Yet, to the best of our knowledge, their charge-noise performance has not been systematically characterized.

In this work, we address both of these two challenges by characterizing QDs fabricated on an ultra-shallow Ge/SiGe heterostructure with a $d \approx \SI{4}{\nano \meter}$ SiGe spacer and using low-temperature ($\sim \SI{100}{\celsius} - \SI{150}{\celsius}$) ALD oxides. We demonstrate that the studied platform can be used for stable QDs, and the electrostatic noise assessment reveals noise levels comparable to those on shallow cap ($\sim \SI{20}{\nano \meter}$) wafers and with high-temperature ALD oxides.

We fabricated two types of devices, dubbed Device A and Device B. Both are fabricated on the same wafer, with the SiGe cap grown to be within $\SI{3}{\nano \meter} - \SI{5.5}{\nano\meter}$. The Atomic resolution high-angle annular dark field (HAADF) - scanning transmission electron microscopy (STEM) image is shown in \cref{fig:figure1}a. Details on the wafer growth and its properties can be found in Ref.\onlinecite{valentini_parity-conserving_2024}. Device A is a double quantum dot (DQD) with a charge sensor; the scanning electron microscopy (SEM) image of it is shown in \cref{fig:figure1}b. Device B is a single QD, see SEM and transport data in supplementary information (SI), Sec.~2.  As a comparison, we also measured transport through the charge sensor used in the spin qubit experiment of Ref.\onlinecite{saez-mollejo_exchange_2025} (dubbed Device C), where the SiGe cap is $d \approx \SI{20}{\nano \meter}$.

Devices A and B have the same fabrication stack except for the ALD-grown oxide: Device A has HfO$_2$ grown at $\SI{150}{\celsius}$, while Device B has Al$_2$O$_3$ grown at $\SI{100}{\celsius}$. Details on the fabrication and layout of Device C are given in Ref.\onlinecite{saez-mollejo_exchange_2025}. The oxide recipes for devices A and B are adapted from the high-temperature ones used in Refs.\onlinecite{jirovec_singlet-triplet_2021, saez-mollejo_exchange_2025} (Device C) to the low-temperature deposition by extending the deposition times. We observed that extending deposition times is crucially important to counteract low gas kinetics due to the reduced temperature (see SI, Sec.~1 for the fabrication details and discussion within).

\begin{figure}[t]
    \centering
  \includegraphics[width=\columnwidth]{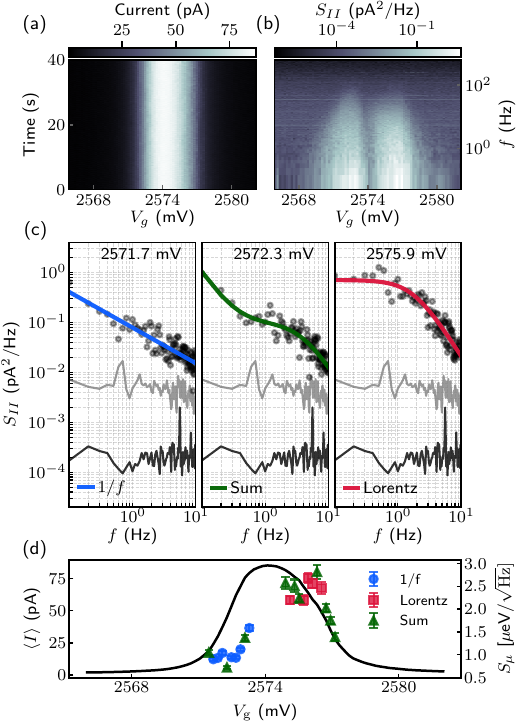}
  \caption{Noise analysis across a Coulomb peak. (a) Time traces of current across the Coulomb peak. The remarkable stability is demonstrated within the $\approx1$ hour data recording window. The external voltage bias $V_{\textrm{bias}}$ is set to $\SI{270}{\micro \volt}$. The voltage drop $V_{\mathrm{SD}}$ over the QD changes by $\approx 30\%$ due to the TIA input resistance $R_{\mathrm{in}} \approx \SI{1}{\mega\ohm}$. (b) PSD estimated via the Welch method for each of the plunger gate voltages $V_{\mathrm{g}}$. The horizontal lines, independent of the plunger gate voltage $V_{\mathrm{g}}$, are attributed either to mechanical vibrations or to the $\SI{50}{\hertz}$ harmonics. (c) Traces demonstrating different spectral behavior (black dots) with the corresponding fit (blue, green, red curves). The corresponding voltage $V_{\mathrm{g}}$ at which the traces are taken are indicated at the top. For all subpanels we plot the PSD in the Coulomb blockade (black) and on top of the Coulomb peak (gray). (d) Averaged current $\left<I(V_{\mathrm{g}})\right>$ (left axis) and electrochemical potential noise density $S_{\mu}$ (right axis) evaluated at $\SI{1}{\hertz}$. The colorcoding of the points is the same as in panel c. The error bars correspond to the fitting errors.}
  \label{fig:figure2}
\end{figure}

We showcase full functionality of Device A in \cref{fig:figure1}c-f. In \cref{fig:figure1}c the stability diagram of the left dot are presented; similar measurements for the right and sensor dots can be found in SI, Sec.~2. For all the QDs we find the charging energy to be within the range $E_{\mathrm{C}} \sim \SI{1.5}{\milli \electronvolt} - \SI{3}{\milli \electronvolt}$, similar to the QDs fabricated on thicker cap wafers\cite{jirovec_singlet-triplet_2021, saez-mollejo_exchange_2025}. The lowest excited states are found to have orbital energies $E_{\textrm{orb}} \sim \SI{0.2}{\milli \electronvolt} - \SI{0.3}{\milli \electronvolt}$. In \cref{fig:figure1}d we present a DQD stability diagram measured via the current response of the charge sensor dot. As shown in \cref{fig:figure1}d, the transition lines for both dots are roughly equidistant and correspond to the addition energy equal to the charging energy $E_{\textrm{add}} \approx E_C$, which implies that the effect of level quantization is not visible yet. We attribute it to the high charge density in both dots. Lithographic squeezing of the dots, surface treatment\cite{sangwan_impact_2025}, control of the fixed oxide charge\cite{simonControlFixedCharge2015} or multi-layer gate structures\cite{zajacReconfigurableGateArchitecture2015} would be further required to reach the few-hole regime. Nevertheless, we demonstrate full charge control via fast gates in \cref{fig:figure1}e-f, where, following Ref.\onlinecite{hendrickx_single-hole_2020}, we pulse through the DQD transition across the virtual detuning and energy axes. We observe charge triangles, which we attribute to charge latching rather than Pauli spin blockade, since the effect appears on both sides of the charge transition (although it is more pronounced on one side).

We now turn to assessing the electrostatic noise levels for Devices A, B, and C. It is commonly argued that the dominant source of low-frequency charge noise stems from two-level fluctuators (TLFs)\cite{duttaLowfrequencyFluctuationsSolids1981}, which leads to the spin-qubits' decoherence\cite{burkard_semiconductor_2023} and reduced two-qubit gate fidelities\cite{shehataModelingSemiconductorSpin2023}. In order to characterize the noise performance of the ultra-shallow platform, we use the widely accepted metric: the electrochemical potential noise density $S_{\mu}$ at $\SI{1}{\hertz}$\cite{lodariLowPercolationDensity2021b, holman3DIntegrationMeasurement2021b, connorsLowfrequencyChargeNoise2019a, elsayedLowChargeNoise2024a, stehouwerExploitingStrainedEpitaxial2025a}.

We estimate the noise levels using the so-called flank method\cite{lodariLowPercolationDensity2021b, holmanCircuitQuantumElectrodynamics, spenceProbingLowFrequencyCharge2023a, chanrionChargeDetectionArray2020}. The detailed description of data acquisition and analysis can be found in the SI, Sec.~3; here we outline the main steps of the procedure. First, we collect time traces of the current $I(t)$ through the QD across a Coulomb peak at a constant external bias voltage $V_{\mathrm{bias}}$. In \cref{fig:figure2}a, we plot time traces corresponding to the 3rd Coulomb resonance of the dataset presented in \cref{fig:figure1}c (Device A). The voltage drop over the quantum dot $V_{\mathrm{SD}} = V_{\mathrm{bias}} - IR_{\mathrm{in}}$ is not kept constant due to the input resistance $R_{\mathrm{in}} \approx \SI{1}{\mega\ohm}$ of our transimpedance amplifier (TIA).  Some datasets were additionally measured with $R_{\mathrm{in}} \approx \SI{100}{\kilo\ohm}$ with no qualitative difference found (see SI, Sec.~6 for further details). 

The PSD traces $S_{II}(f, V_{\mathrm{g}})$ for each plunger gate voltage $V_{\mathrm{g}}$, estimated using the Welch method, are shown as a 2D map in \cref{fig:figure2}b. A pronounced dip in low-frequency noise power is observed at the top of the Coulomb peak ($V_{\mathrm{g}} = \SI{2574}{\milli\volt}$), where the transconductance $\partial I/\partial V_{\mathrm{g}}$ approaches zero. Since the contribution of the electrochemical potential fluctuations $\delta \mu(t)$ to the current fluctuations $\delta I(t)$ vanishes at zero transconductance, the noise spectrum -- both at the top of the peak and deep in the Coulomb blockade -- is expected to be dominated by the noise added by the TIA\cite{lodariLowPercolationDensity2021b, holmanCircuitQuantumElectrodynamics}. The substantial difference between the noise levels measured at the peak top and in blockade (black and gray traces in \cref{fig:figure2}c) reflects the strong dependence of the TIA-added noise on the load impedance seen by the amplifier. A rigorous subtraction of the TIA noise would therefore require independent measurements of the device differential conductance, as well as the noise of the TIA for that specific load impedance. Such a procedure is beyond the scope of this work, and we therefore do not subtract the TIA-added noise explicitly. Nevertheless, for the noise levels measured in our devices, quantitative conclusions can be drawn without this detailed characterization, as we discuss below.

While the standard TIA noise model includes equivalent input current and voltage noise sources, as well as Johnson-Nyquist noise from the feedback resistor\cite{kretinin_wide-band_2012}, in our TIA, there is an additional contribution associated with the input impedance of the amplifier $R_{\mathrm{in}} \approx \SI{1}{\mega\ohm}$ (or $R_{\mathrm{in}} \approx \SI{100}{\kilo\ohm}$, depending on the configuration). As a result, the total input-referred current noise not only increases monotonically as the load impedance decreases, but also saturates at its largest value when the device resistance is minimal and the load impedance is dominated by $R_{\mathrm{in}}$ (see SI, Sec.~4). We observe that the measured noise at the top of the Coulomb peak approaches the maximal TIA-added noise and use this spectrum as an empirical reference to select spectra acquired on the flanks for further analysis. More concretely, only those spectra whose noise level exceeds the peak-top reference in the frequency range from $\SI{100}{\milli \hertz}$ to $\SI{10}{\hertz}$ are retained. The frequency window is chosen for the subsequent fits such that it spans around the frequency of interest of $\SI{1}{\hertz}$, while remaining well separated from the TIA roll-off at higher frequencies ($\approx \SI{30}{\hertz}$). Nearly all spectra selected in this way also exceed the maximal saturate TIA noise at $\SI{1}{\hertz}$, indicating that the extracted noise is predominantly of the device origin. 

A large fraction of the selected PSD trends $S_{II}$ cannot be fitted purely by the $1/f^{\beta}$ function. This is expected if the contribution of a single TLF is dominant over the distribution of TLFs\cite{shehataModelingSemiconductorSpin2023} and has already been observed in Refs.\onlinecite{chanrionChargeDetectionArray2020, elsayedLowChargeNoise2024a}. Following Ref.\onlinecite{elsayedLowChargeNoise2024a}, we define three models:
\begin{equation}
    S_{1/f}(f) = \frac{A}{f^{\beta}},
    \label{eq:1f}
\end{equation}
\begin{equation}
    S_{\textrm{Lorentz}}(f) = \frac{B}{(f/f_0)^2 + 1},
    \label{eq:lorentz_model}
\end{equation}
\begin{equation}
    S_{\textrm{Sum}}(f) = \frac{A}{f^{\beta}} + \frac{B}{(f/f_0)^2 + 1},
    \label{eq:sum_model}
\end{equation}
where $A$, $B$, $\beta$, $f_0$ are the fitting parameters. \cref{eq:1f} describes the spectrum of the TLF distribution, \cref{eq:lorentz_model} a single TLF, and \cref{eq:sum_model} captures the contribution of both. We fit all three models and select the one that minimizes the Bayesian Information Criterion (BIC), which penalizes models with more fitting parameters. In \cref{fig:figure2}c examples of the fits for three values of $V_{\mathrm{g}}$ are demonstrated. A clear distinction in the PSD trends is observed.

The transconductance at fixed external bias voltage $\left(\partial I/\partial V_{\mathrm{g}}\right)_{V_{\mathrm{bias}}}$ is extracted by taking the numerical derivative of the averaged current $\left<I\right>$, while the lever-arm $\alpha$ is obtained from the slopes of the corresponding Coulomb diamond. The lever arms vary between $0.07-\SI{0.11}{\electronvolt}/\SI{}{\volt}$ for Devices A and B.
The PSD of the electrochemical potential fluctuations is calculated as (see SI, Sec.~4 for the derivation):
\begin{equation}
    S_{\mu\mu}(f) = \alpha^{2}\, S_{VV}(f) = \alpha^{2}\left(\frac{\partial I}{\partial V_{\mathrm{g}}}\right)_{V_{\mathrm{bias}}}^{-2} S_{II}(f).
    \label{eq:mu_noise}
\end{equation}

\begin{figure}[t]
    \centering
  \includegraphics[width=\columnwidth]{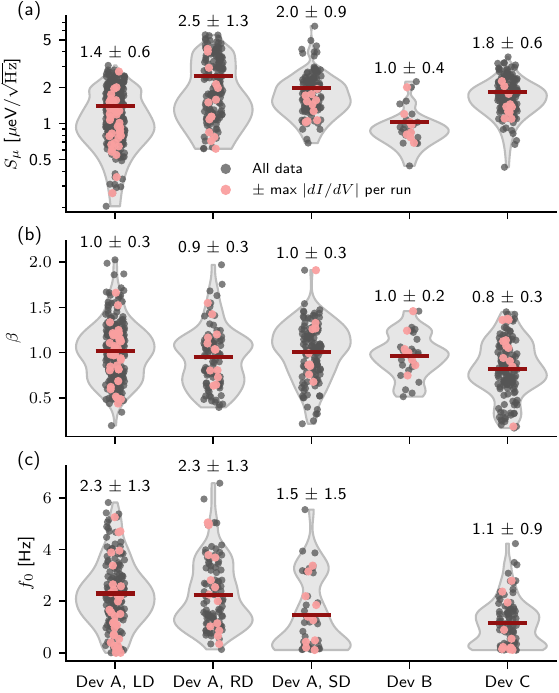}
  \caption{Statistics of the noise metrics. The value on top of each of the violin plots corresponds to the average and standard deviation values for all data for that specific device and is marked by the bar. (a) Charge noise value evaluated at $\SI{1}{\hertz}$. (b) Extracted exponent $\beta$ reported for those datapoints described by the model with the $1/f$ term. (c) Extracted frequency $f_0$ of the single TLF. No Lorentz contributions were observed for Device B.}
  \label{fig:figure3}
\end{figure}

 In \cref{fig:figure2}d we plot the values of noise $S_{\mu}$ at $\SI{1}{\hertz}$, where $S_{\mu} = \sqrt{S_{\mu\mu}}$, which are calculated from the fitted models throughout the Coulomb peak for selected gate voltages $[V_{\mathrm{g}}]$. We highlight the PSD-model by color coding: $1/f$ -- blue, Lorentz -- red and Sum -- green. Theoretically, the extracted noise amplitude $S_{\mu}(\SI{1}{\hertz})$ should be independent of $V_{\mathrm{g}}$ (see toy-model simulations in the SI, Sec.~5). Instead, we observe a clear non-uniform distribution across the Coulomb peak. The noise differs between the two flanks of the Coulomb peak, both in magnitude and in its spectral character: on the left it is consistent with an ensemble of TLFs, whereas on the right it is dominated by a single prominent TLF. One possible explanation is that changing $V_{\mathrm{g}}$ and $V_{\mathrm{SD}}$ rearranges local screening charges and trap occupancies, thereby modifying the coupling to individual fluctuators. This non-uniformity of $S_{\mu}$ is not unique to our data: it was observed in Refs.\onlinecite{spenceProbingLowFrequencyCharge2023a, chanrionChargeDetectionArray2020} and studied rigorously in Ref.\onlinecite{ye_characterization_2024}.

Following the protocol described above, we collected datasets for devices A, B, and C. For Device A, the noise on the Sensor QD is measured when the DQD gates are energized; on the Left (Right) QD when the Sensor QD gates are energized and those of the Right (Left) are at ground potential. All the analyzed datasets can be found in the SI, Sec.~6. We report the electrochemical potential noise density $S_{\mu}$ at $\SI{1}{\hertz}$ (\cref{fig:figure3}a), $\beta$ for the PSD traces exhibiting $1/f$-trend (\cref{fig:figure3}b), and $f_0$ for the Lorentzians (\cref{fig:figure3}c). Given that most of the charge noise studies\cite{lodariLowPercolationDensity2021b, elsayedLowChargeNoise2024a, stehouwerExploitingStrainedEpitaxial2025a} use the noise evaluated at points with maximal transconductance, we highlight these operating points by pink color. The most important feature of \cref{fig:figure3}a is the fact that both devices A and B exhibit noise statistics comparable to that of Device C (the sensor dot in Ref.\onlinecite{saez-mollejo_exchange_2025}). The latter has demonstrated a microwave driven qubit with dephasing times up to $T_2^* = \SI{3}{\micro \second}$. We interpret this comparison as a promising result for the ultra-shallow heterostructures. Additionally, the exponents $\beta$ are distributed around the average value of 1, which is predicted by theoretical modeling\cite{shehataModelingSemiconductorSpin2023}. The switching frequencies $f_0$ are distributed in the low $\SI{}{\hertz}$ range; we did not observe any Lorentz behavior for Device B.

\begin{table}[t]
\centering
\tiny
\setlength{\tabcolsep}{2.5pt}
\renewcommand{\arraystretch}{1.05}

\begin{threeparttable}
\caption{Selection of reported charge-noise amplitudes $S_{\mu}$ (in $\mu$eV/$\sqrt{\mathrm{Hz}}$) at 1 Hz across various material platforms with different spacer thicknesses $d$ ($\SI{}{\nano \meter}$) measured via the flank method. If a reference reports different numbers upon improving the fabrication flow, we take the best result of that work.}
\label{tab:noise_lit}

\begin{tabularx}{\columnwidth}{
>{\raggedright\arraybackslash}p{0.32\columnwidth}
>{\raggedright\arraybackslash}p{0.3\columnwidth}
>{\raggedleft\arraybackslash}p{0.1\columnwidth}
>{\raggedleft\arraybackslash}p{0.15\columnwidth}}
\toprule
Reference & Material & d & $S_{\mu}$ \\
\specialrule{1.2pt}{0pt}{0pt}

Kim et al.\cite{kimLowdisorderMetaloxidesiliconDouble2019} & Si/SiO$_2$ & - & 3.4 \\
Petit et al.\cite{petitSpinLifetimeCharge2018} & ${}^{28}$Si/SiO$_2$  & - & 2 \\
Chanrion et al.\cite{chanrionChargeDetectionArray2020} & Si/SiO$_2$ & - & 1 \\
Elsayed et al.\cite{elsayedLowChargeNoise2024a} & Si/SiO$_2$ & - & 0.61 \\
Freeman et al.\cite{freemanComparisonLowFrequency2016} & Si/SiO$_2$ & - & 0.49 \\

\specialrule{1.2pt}{0pt}{0pt}
Spence et al.\cite{spenceProbingLowFrequencyCharge2023a} &  CMOS NW& - & 1.73 \\
Jekat et al.\cite{jekatExfoliatedHexagonalBN2020a} & InSb NW& - & 1 \\
\specialrule{1.2pt}{0pt}{0pt}
Freeman et al.\cite{freemanComparisonLowFrequency2016} & Si/SiGe & 40 & 2 \\
Connors et al.\cite{connorsLowfrequencyChargeNoise2019a} & Si/SiGe & 50& 0.84 \\
Holman et al.\cite{holman3DIntegrationMeasurement2021b} & ${}^{28}$Si/SiGe & 42 & 0.77 \\
Struck et al.\cite{struckLowfrequencySpinQubit2020a} & ${}^{28}$Si/SiGe & 45 & 0.47 \\
Mi et al.\cite{miLandauZenerInterferometryValleyorbit2018} & Si/SiGe & 50 & 0.33 \\
Paquelet et al.\cite{paqueletwuetzReducingChargeNoise2023a} & ${}^{28}$Si/SiGe & 30 & 0.29 \\
\specialrule{1.2pt}{0pt}{0pt}
This work & Ge/SiGe & 4 & 1.8 \\
 & Ge/SiGe & 20 & 1.8 \\
Hendrickx et al.\cite{hendrickxGatecontrolledQuantumDots2018a} & Ge/SiGe & 22 & 1.4 \\
Lodari et al.\cite{lodariLowPercolationDensity2021b} & Ge/SiGe & 55 & 0.6 \\
Stehouwer et al.\cite{stehouwerExploitingStrainedEpitaxial2025a} & Ge/SiGe & 55 & 0.3 \\
\bottomrule
\end{tabularx}
\end{threeparttable}
\end{table}

To conclude, we relate the average noise level observed for Devices A and B to those reported in the literature (see \cref{tab:noise_lit}). We find it comparable to most of the shallow cap wafers and CMOS structures. We note, however, that it is almost six times worse than the best Ge/SiGe heterostructure with the thick $\SI{55}{\nano\meter}$ cap. Although the latter is undoubtedly preferred for quantum computing applications, ultra-shallow cap wafers offer certain advantages for prototyping hybrid devices. In principle, any material can be deposited or placed in proximity to the Ge layer upon removal of the native SiO$_2$ layer. This concerns not only the superconducting materials but also ferromagnetic (\eg EuS) or two-dimensional materials with enhanced spin-orbit interaction such as WSe$_2$. Finally, a tunable proximitized QD was very recently formed on such an ultra-shallow Ge/SiGe heterostructure, demonstrating its potential for future hybrid spin-superconductor experiments\cite{fabrisGranularAluminumInduced2026}.

\begin{acknowledgments}
We sincerely thank Nick van Loo, Greg Mazur, Dhananjay Joshi and Srijit Goswami for their inputs on low-temperature HfOx deposition, Matias Urdampilleta and Daniel Jirovec for the discussions, and Kristen L\'{e}onard for the careful read of the manuscript. This research was supported by the Scientific Service Units of ISTA through resources provided by the MIBA Machine Shop and the Nanofabrication facility. The authors acknowledge support from the NOMIS Foundation, the European Innovation Council Pathfinder grant no. 101115315 (QuKiT), the FWF Projects with DOI:10.55776/F86,  DOI:10.55776/PAT7682124, and DOI:10.55776/P36507, and the HE-MSCA-PF project with DOI:10.3030/101150858. ICN2 is supported by the Severo Ochoa program from Spanish MCIN / AEI (Grant No.: CEX2021-001214-S) and is funded by the CERCA Program / Generalitat de Catalunya. ICN2 acknowledges funding from Generalitat de Catalunya 2021SGR00457. We acknowledge support from CSIC Interdisciplinary Thematic Platform (PTI+) on Quantum Technologies (PTI-QTEP+).

\end{acknowledgments}

\section*{Author Declarations}
\subsection*{Conflict of Interest}
The authors have no conflicts to disclose.

\subsection*{Author contributions}
\textbf{Maksim Borovkov}: Conceptualization (equal); Formal analysis (lead);
Investigation (equal); Methodology (lead); Resources (equal); Validation (equal);
Visualization (lead); Writing – original draft (lead); Writing –
review \& editing (equal). 
\textbf{Yona Schell}: Investigation (equal); Writing – original draft (supporting).
\textbf{Dina Sokolova}: Resources (equal).
\textbf{Kevin Roux}: Investigation (supporting); Resources (supporting).
\textbf{Paul Falthansl-Scheinecker}: Investigation (supporting); Resources (equal).
\textbf{Giorgio Fabris}: Resources (equal).
\textbf{Devashish Shah}: Investigation (supporting).
\textbf{Jaime Saez-Mollejo}: Investigation (supporting).
\textbf{Rodolfo Previdi}: Resources (supporting).
\textbf{Inas Taha}: Investigation (supporting).
\textbf{Aziz Gen\c{c}}: Investigation (supporting).
\textbf{Jordi Arbiol}: Investigation (supporting).
\textbf{Stefano Calcaterra}: Resources (equal).
\textbf{Afonso De Cerdeira Oliveira}: Resources (equal).
\textbf{Daniel Chrastina}: Resources (equal).
\textbf{Giovanni Isella}: Resources (equal).
\textbf{Anton Bubis}: Investigation (supporting); Supervision (supporting); Resources (equal); Writing –
review \& editing (equal).
\textbf{Georgios Katsaros}: Conceptualization (equal); Supervision (lead); Funding acquisition (lead); Validation (equal); Project administration (lead); Writing – original draft (supporting); Writing -- review \& editing (equal).

\section*{Data availability}
All data included in this work will be available at the
Institute of Science and Technology Austria repository.

\section*{References}

\bibliographystyle{apsrev4-2}
\bibliography{main_references2.bib}

\appendix

\setcounter{section}{0}
\renewcommand{\thesection}{S\arabic{section}}
\renewcommand{\thesubsection}{S\arabic{section}.\arabic{subsection}}
\setcounter{figure}{0}
\renewcommand{\thefigure}{S\arabic{figure}}
\setcounter{table}{0}
\renewcommand{\thetable}{S\arabic{table}}
\setcounter{equation}{0}
\renewcommand{\theequation}{S\arabic{equation}}
\renewcommand{\appendixname}{}
\onecolumngrid
\newpage

\begin{center}

\noindent\begin{minipage}{\textwidth}
{\Large\bfseries Supplementary Information for:}\\[4pt]
{\large\bfseries 
Low-Noise Quantum Dots in Ultra-Shallow Ge/SiGe\\
Heterostructures for Prototyping Hybrid Semiconducting-Superconducting Devices}
\end{minipage}\\[8pt]
\end{center}

\section{Fabrication details.}

In Table \ref{tab:fabrication_summary_ABC} we summarize fabrication recipes for Devices A-C. For Devices A and B, the ALD oxide was grown in the Oxford FlexlAL machine at reduced temperatures: $\SI{150}{\celsius}$ and $\SI{100}{\celsius}$, respectively. Following Refs.\onlinecite{georgeAtomicLayerDeposition2010, shekhar_low-temperature_2022, paghiCryogenicBehaviorHighPermittivity2024}, we adjusted the dosing and purging times to compensate for reduced gas kinetics. The low-temperature recipes were not further optimized, and systematic optimization is left for future work. 

\newlength{\FabLeftColW}
\newlength{\FabDevColW}

\setlength{\FabLeftColW}{0.1\textwidth} 
\setlength{\FabDevColW}{0.25\textwidth}

\begin{table*}[h!] 
\centering
\small
\setlength{\tabcolsep}{10pt}
\renewcommand{\arraystretch}{2.5}

\label{tab:fabrication_summary_ABC}

\begin{tabular}{|>{\bfseries}p{\FabLeftColW}|
                >{\centering\arraybackslash}p{\FabDevColW}|
                >{\centering\arraybackslash}p{\FabDevColW}|
                >{\centering\arraybackslash}p{\FabDevColW}|}
\hline
\bfseries & \bfseries Device A & \bfseries Device B & \bfseries Device C\cite{saez-mollejo_exchange_2025} \\
\hline
Contacts & Ar milling + e-beam Pt evaporation & HF dip + e-beam Pd evaporation & Ar milling + e-beam Pt evaporation \\
\hline
Mesa &
\multicolumn{3}{>{\centering\arraybackslash}p{\dimexpr 3\FabDevColW + 4\tabcolsep + 2\arrayrulewidth\relax}|}{Reactive Ion Etching in SF$_6$/CHF$_3$/O$_2$ plasma} \\
\hline
ALD &
{\raggedright\renewcommand{\arraystretch}{1.2}\begin{tabular}[t]{@{}l@{}}
\textbf{HfO$_2$ @150C}\\
Precursor: TEMAH\\
TEMAH \textit{Dose 1\,s}\\
TEMAH\textit{Purge 50\,s}\\
H$_2$O \textit{Dose 50\,ms}\\
H$_2$O \textit{Purge 50\,s}
\end{tabular}}
&
{\raggedright\renewcommand{\arraystretch}{1.2}\begin{tabular}[t]{@{}l@{}}
\textbf{Al$_2$O$_3$ @100C}\\
Precursor: TMA\\
TMA \textit{Dose 1\,s}\\
TMA \textit{Purge 120\,s}\\
H$_2$O \textit{Dose 1\,s}\\
H$_2$O \textit{Purge 120\,s}\\
\\
\end{tabular}}
&
{\raggedright\renewcommand{\arraystretch}{1.2}\begin{tabular}[t]{@{}l@{}}
\textbf{HF dip + Al$_2$O$_3$ @300C}\\
Precursor: TMA\\
TMA \textit{Dose 20\,ms}\\
TMA \textit{Purge 2000\,ms}\\
H$_2$O \textit{Dose 100\,ms}\\
H$_2$O \textit{Purge 5000\,ms}
\end{tabular}}
\\
\hline
Gates &
\multicolumn{3}{>{\centering\arraybackslash}p{\dimexpr 3\FabDevColW + 4\tabcolsep + 2\arrayrulewidth\relax}|}{E-beam lithography; Ti (3\,nm) / Pd (20--30\,nm) e-beam deposition} \\
\hline
\end{tabular}

\caption{Fabrication process summary for Devices A--C. Mesa etching and gate writing are identical across devices, while ohmic-contact metallization and ALD dielectric deposition follow device-specific recipes.}
\end{table*}

For the HAADF-STEM image presented in the main text an electron-transparent lamella was prepared using a Thermo Fisher Scientific (TFS) Helios 5 UX focused-ion beam (FIB). Aberration-corrected scanning transmission electron microscopy (AC-STEM) was performed using a probe and image corrected TFS Spectra 300 operating at 300 kV. HAADF images were collected using a detector inner collection angle of 63 mrad and an outer angle of 200 mrad. The imaging was performed at a probe current of 100 pA and convergence angle of 20 mrad.

\newpage
\section{Additional Stability diagrams.}
\begin{figure*}[h!]
  \centering
  \includegraphics[width=\textwidth]{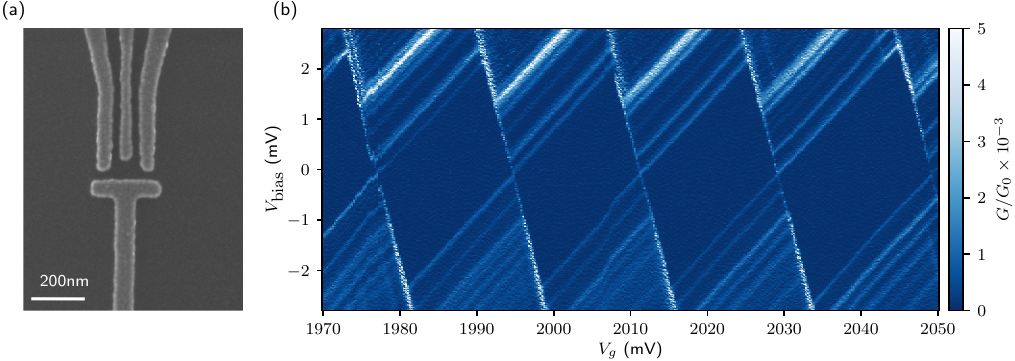}
  \caption{Device B overview. (a) SEM image of Device B. (b) Stability diagram of Device B.}
  \label{fig:figureS1}
\end{figure*}

\begin{figure*}[h!]
  \centering
  \includegraphics[width=\textwidth]{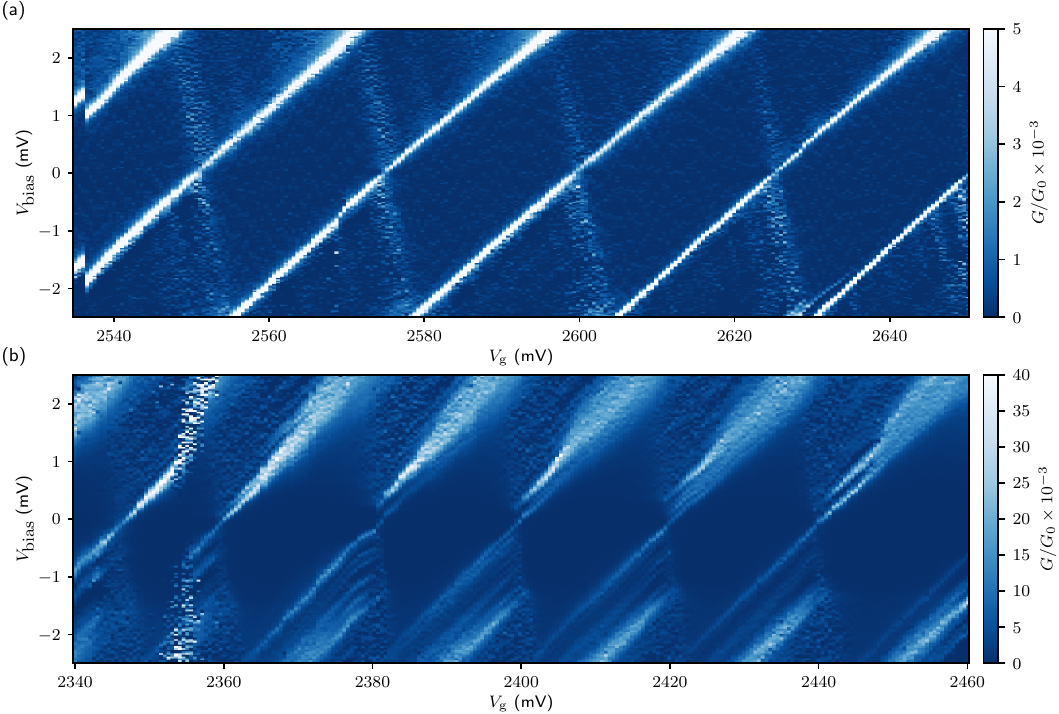}
  \caption{Stability diagrams for (a) Right QD of Device A. (b) Sensor QD of Device A. The Sensor QD was tuned into the cotunneling regime for charge sensing. The noise on the Sensor QD was analyzed in that gate configuration.}
  \label{fig:figureS1}
\end{figure*}

\newpage
\section{Noise estimation procedure.}\label{app:noise_analysis}
We record $\SI{40}{\second}$-long current time traces $I(t)$ through the QD at equally spaced values of the plunger gate voltage $V_{\mathrm{g}}$ as it steps across a Coulomb peak. The time traces are acquired using a \textit{Keysight 34465A} digital multimeter at a sampling rate of $f_s = \SI{1.25}{\kilo \hertz}$. We estimate the current PSD $S_{II}$ using the Welch's periodogram method with a Hann window and $50\%$ overlap\cite{welch_use_1967}. Within this method, the time trace is divided into $N_{\textrm{seg}}$ segments which are half-overlapped, and the corresponding windowed periodograms are then averaged. The number $N_{\textrm{seg}}$ of segments is chosen empirically as a compromise: larger $N_{\textrm{seg}}$ decreases the dispersion of the PSD estimation at the cost of increasing the minimal frequency $f_{\mathrm{min}} = 1/T_{\textrm{seg}} =  f_sN/N_{\textrm{seg}}$. We found that the PSD evaluated with $N_{\textrm{seg}}=2$ had too large a dispersion, while $N_{\textrm{seg}}=8$ had too large a minimal frequency $f_{\mathrm{min}}$ for a robust fit of the spectra. As such, we landed at $N_{\mathrm{seg}} = 4$. The Hann window is used to suppress spectral leakage, which in our case would otherwise transfer power from the strong low-frequency $\sim 1/f$ components into neighboring frequency bins. Ref.\onlinecite{elsayedLowChargeNoise2024a} is an example of a study where the current PSD $S_{II}$ was estimated using Hann window.

From the time traces we compute the mean current $\langle I(V_g)\rangle$ and its transconductance $d\langle I\rangle/dV_g$ (after interpolation to a twice-denser grid). We preselect points in $V_{\mathrm{g}}$ at which $dI/dV_{\mathrm{g}}$ exceeds $20\%$ of the maximal value of the derivative, thereby excluding gate voltages $V_{\mathrm{g}}$ in the regions of the Coulomb peak apex and Coulomb blockade. We then retain only those spectra that can be distinguished from the PSD correspondent to the top of the Coulomb peak in the frequency range $(f_{\mathrm{min}}, f_{\mathrm{cutoff}})$. The cutoff frequency is set to $\SI{10}{\hertz}$ to avoid the roll-off effects of the TIA and $\SI{50}{\hertz}$ harmonics. We then define $f^*$ as the frequency above which the PSD is no longer consistently above or below $S_{\mathrm{top}}$ and fluctuates around it. Practically, $f^*$ is found by a sliding window, where $\log\!\big[S(f)/S_{\mathrm{top}}(f)\big]$ has roughly equal numbers of positive and negative bins. We then take PSDs for which $f^* > f_{\mathrm{cutoff}}$ holds and ensure that it is above $S_{\mathrm{top}}$ within the frequency band. 

We noticed that the major part of our PSD does not follow the $1/f$-trend precisely. As such, we decided to use different models to fit the data. To choose the best fit algorithmically, we compute the Bayesian Information Criterion $\textrm{BIC} = n \ln{(\textrm{RSS}/n)} + k\ln{(n)}$, where $n$ is the number of datapoints, $k$ is the number of the model fit parameters and $\textrm{RSS} = \frac{1}{n}\sum_i(y_i -\hat{y}_i)^2$ is the residual sum of squares (minimized during the fit). We then pick the fit with the lowest $\textrm{BIC}$. We observed that fitting $\log$ is more robust, as it accounts for the uncertainty in least-squares multiplicatively, effectively treating different decades of the PSD equally. The $\textrm{BIC}$ can be understood as a parameter that is minimal for the best log-likelihood function, which is also penalizing the $\textrm{Sum}$-model for extra fitting parameters. Empirically, we observed, that sometimes even when $\textrm{BIC}$ of the $\textrm{Sum}$-model is winning, the dispersion on its parameters are larger than its values. In this case, we drop it and choose the second best in $\textrm{BIC}$ model.

\begin{figure}[b!]
    \centering
    \includegraphics[width=\textwidth]{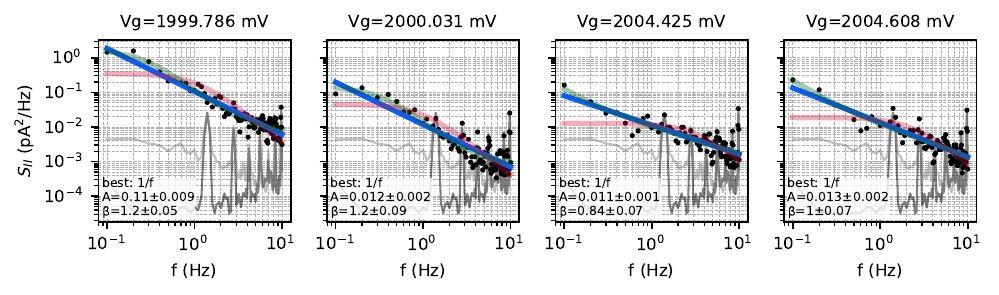}
    \caption{Fitting examples for Device B. The background noise is more pronounced for Device B. One can see significant pulse-tube vibrations component at $\SI{1.4}{\hertz}$. It is important to note that we do not remove peak points for the fit and rely on the fact that their influence is negligible.}
    \label{fig:devB_noise_fits}
\end{figure}

\begin{figure}[t!]
    \centering
    \includegraphics[width=\textwidth]{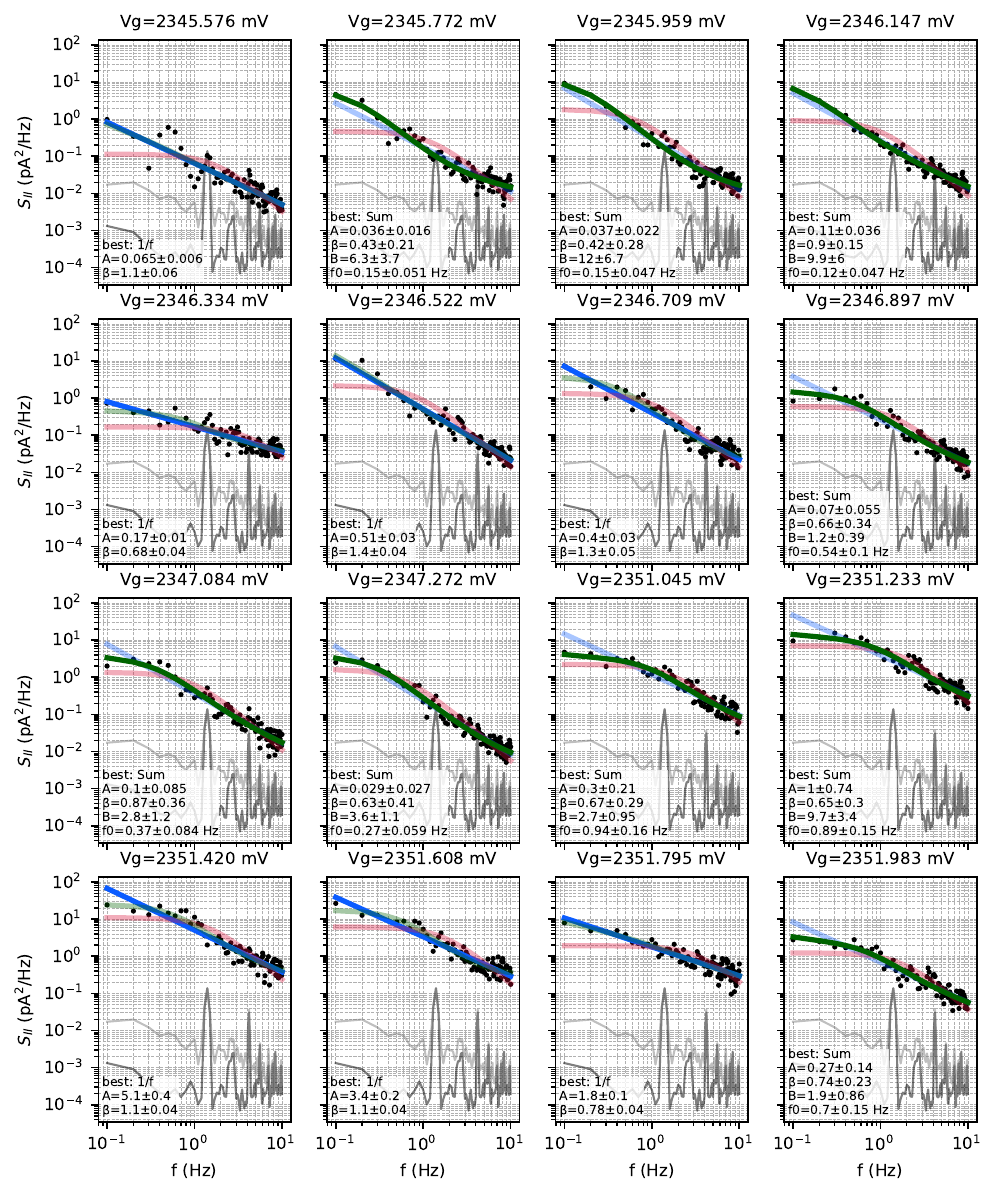}
    \caption{Fitting examples for Device C.}
    \label{fig:devC_noise_fits}
\end{figure}

Examples of fits for Device A are given in the main text; some examples of the fits over selected voltages for Devices B and C are shown in \cref{fig:devB_noise_fits,fig:devC_noise_fits}. In both \cref{fig:devB_noise_fits,fig:devC_noise_fits}, the light gray trace represents PSD at the top of the Coulomb peak and the dark gray in Coulomb blockade, at low frequencies they are $1-2$ orders of magnitude different from each other.

After the fit and its parameters are obtained, we compute $S_{II}(\SI{1}{\hertz})$ and, using Eq.~4 of the main text, we obtain $S_{\mu}$. The distribution of $S_{\mu}$ among the devices and analyzed Coulomb peaks is summarized in \cref{fig:SI_summary_coulomb}

\newpage
\section{Impact of the amplifier input resistance $R_{\mathrm{in}}$ on the noise estimation.}\label{sec:theory}

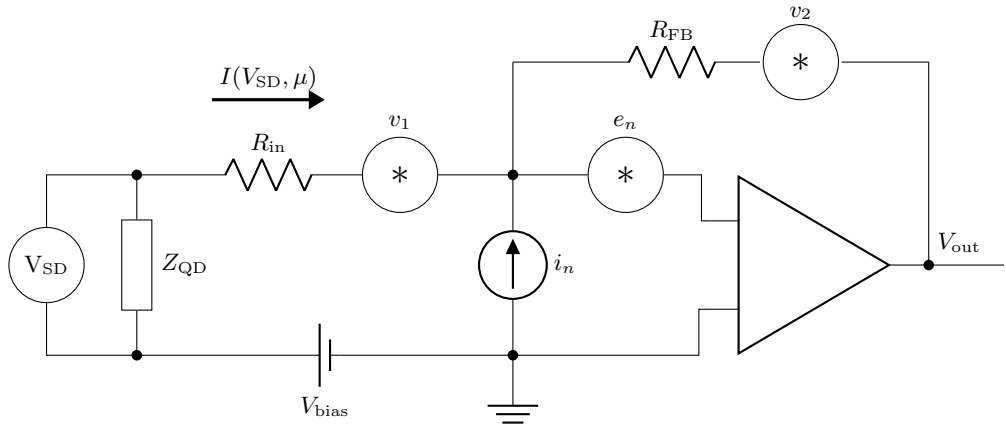
\begin{figure}[b!] 
\centering
\begin{circuitikz}[american]
\tikzset{
  circ10/.style={circle,draw,line width=1.2pt,minimum size=10mm,inner sep=0pt},
  qdblk/.style={draw,line width=1.2pt,minimum width=4mm,minimum height=12mm,inner sep=0pt}
}

\coordinate (A) at (0,0);      
\fill (A) circle (2pt);
\coordinate (B) at (0,-2.4);     
\fill (B) circle (2pt);
\coordinate (G) at (5,-2.4);     
\coordinate (INBOT) at (5,-2.4);
\fill (INBOT) circle (2pt);

\draw (A) -- ++(0.0,0) coordinate (QDtop);
\draw (B) -- ++(0.0,0) coordinate (QDbot);

\node[draw, minimum width=4mm, minimum height=12mm, inner sep=0pt] (QD)
    at ($(QDtop)!0.5!(QDbot)$) {};

\node[anchor=west] at (QD.east) {$Z_{\mathrm{QD}}$};
\draw (QDtop) -- (QD.north);
\draw (QDbot) -- (QD.south);

\coordinate (vsd_left) at ($(A)+(-1.2,0)$);

\coordinate (vsd_center) at ($(vsd_left)!0.5!(vsd_left |- B)$);

\node[circle,draw,minimum size=10mm,inner sep=0pt] (VSD) at (vsd_center) {$\mathrm{V_{SD}}$};

\draw (A) -- (vsd_left) -- (VSD.north);
\draw (VSD.south) -- (vsd_left |- B) -- (B);

\draw (B) to[battery1,l_=$V_{\mathrm{bias}}$] (G);
\draw (G) -- (INBOT);

\draw (A) -- (1,0) to[R,l=$R_{\mathrm{in}}$] (2.5,0) -- (3,0);

\node[circle,draw,minimum size=10mm,inner sep=0pt] (Vone) at (3.5,0) {\Large$\ast$};
\node at (3.5,0.7) {$v_1$};
\draw (3,0) -- (Vone.west);

\coordinate (SUM) at (5,0);
\fill (SUM) circle (2pt);
\draw (Vone.east) -- (SUM);

\draw (INBOT) to[I,l_=$i_n$, bipoles/length=15mm] (SUM);

\node[circle,draw,minimum size=10mm,inner sep=0pt] (En) at (6.5,0) {\Large$\ast$};
\node at (6.5,0.7) {$e_n$};
\draw (SUM) -- (En.west);

\node[op amp, noinv input down, scale=1.2] (op) at (9,-1.2) {};

\fill[white] ($(op.+)+(+0.75,0)$) circle (6pt);
\fill[white] ($(op.-)+(+0.75,0)$) circle (6pt);

\draw (En.east) -- ++(0.5,0) |- (op.-);

\draw (op.+) -- ++(-0.1,0) |- (INBOT);

\draw (op.out) -- ++(0.1,0) coordinate (OUT)
      node[right, yshift=3mm] {$V_{\mathrm{out}}$}
      -- ++(1cm,0) coordinate (OUTR);

\coordinate (FBup)  at ($(SUM)+(0,1.5)$);
\coordinate (FBend) at (OUT |- FBup);   

\draw (SUM) -- (FBup);

\coordinate (Rstart) at ($(FBup)+(11mm,0)$);          
\coordinate (V2L)    at ($(FBend)+(-22mm,0)$);

\coordinate (Rend) at ($(V2L)+(-2mm,0)$);

\draw (FBup) -- (Rstart)
      to[R,l=$R_\mathrm{FB}$] (Rend);

\node[circle,draw,minimum size=10mm,inner sep=0pt] (Vtwo) at ($(V2L)+(5mm,0)$) {\Large$\ast$};
\node at (Vtwo.north) [yshift=1.5mm] {$v_2$};

\draw[shorten >=1pt] (Rend) -- (Vtwo.west);
\draw[shorten <=1pt] (Vtwo.east) -- (FBend);

\draw (FBend) |- (OUT);
\fill (OUT) circle (2pt);
\node[ground, scale=1.6] at (INBOT) {};

\draw[->, line width=1.2pt, >=Triangle] ($(A)+(1,+1)$) -- ($(SUM)+(-2.5,+1)$)
  node[midway, above] {$I(V_{\mathrm{SD}}, \mu)$};

\end{circuitikz}
\caption{Simplified model TIA circuit. $V_{\mathrm{SD}}$ represents the voltage drop across the QD. $i_{\mathrm{n}}$, $e_{\mathrm{n}}$ are the current and voltage noise sources of the operational amplifier, respectively. $v_1$ and $v_2$ are the Johnson-Nyquist voltage noise sources for the $R_{\mathrm{in}}$ and $R_{\mathrm{FB}}$ resistors. The current through the QD is $I(V_{\mathrm{SD}}, \mu)$.}
\label{fig:curciut}
\end{figure}

The purpose of this section is twofold. First, we show theoretically how the TIA noise depends on the load impedance and present the experimental characterization of the TIA used in this work. We demonstrate that the presence of the external input resistance \(R_{\mathrm{in}}\) causes the TIA-added noise to saturate at low QD resistances, where the load is dominated by \(R_{\mathrm{in}}\). Second, we derive Eq.~4 of the main text and clarify how the noise estimation formula differs depending on whether the external bias voltage \(V_{\mathrm{bias}}\) or the voltage drop across the QD, \(V_{\mathrm{SD}}\), is held constant. For large \(R_{\mathrm{in}}\), these two situations are not equivalent.

We base our analysis on a simplified TIA model that neglects input capacitances\cite{kretinin_wide-band_2012}. The corresponding circuit is shown in \cref{fig:curciut}. An external bias voltage $V_{\mathrm{bias}}$ is applied across the QD, represented by the impedance $Z_{\mathrm{QD}}$, and the device is connected to the TIA through the input resistance $R_{\mathrm{in}}$ at room temperature. The voltage drop across the QD, $V_{\mathrm{SD}}$, together with the electrochemical potential $\mu$, sets the current $I = I(V_{\mathrm{SD}}, \mu)$. The TIA feedback loop includes a resistor $R_{\mathrm{FB}} = \SI{1}{\giga\ohm}$ (or, for some measurements, $R_{\mathrm{FB}} = \SI{100}{\mega\ohm}$). The output voltage $V_{\mathrm{out}}$ is measured and converted to current as $V_{\mathrm{out}}/R_{\mathrm{FB}}$.

We include four noise sources in the model. The first two, \(v_1\) and \(v_2\), are the Johnson--Nyquist voltage-noise sources associated with the resistors \(R_{\mathrm{in}}\) and \(R_{\mathrm{FB}}\), respectively, both at room temperature. Their spectral densities are
\[
v_1^2 = 4 k_{\mathrm{B}} T R_{\mathrm{in}}, \qquad
v_2^2 = 4 k_{\mathrm{B}} T R_{\mathrm{FB}}.
\]
For the TIA used in this study ~\href{https://qtwork.tudelft.nl/~schouten/ivvi/doc-mod/docm1b.htm}{(\textit{https://qtwork.tudelft.nl/~schouten/ivvi/doc-mod/docm1b.htm})}, \(R_{\mathrm{in}}=\SI{1}{\mega\ohm}\) for \(R_{\mathrm{FB}}=\SI{1}{\giga\ohm}\), while \(R_{\mathrm{in}}=\SI{100}{\kilo\ohm}\) for \(R_{\mathrm{FB}}=\SI{100}{\mega\ohm}\). The other two noise sources are the equivalent input voltage and current noise of the operational amplifier, \(e_N\) and \(i_N\).

The Kirchhoff's laws read
\begin{equation}
\left\{
\begin{aligned}
V_{\mathrm{bias}} - V_{\mathrm{N}} + V_1 &= I\!\left(V_{\mathrm{SD}},\mu\right) R_{\mathrm{in}} + V_{\mathrm{SD}},\\
V_{\mathrm{out}} - V_2 - V_{\mathrm{N}} &= \left(I\!\left(V_{\mathrm{SD}},\mu\right) + I_{\mathrm{N}}\right) R_{\mathrm{FB}} \, ,
\end{aligned} \label{eq:kirch_1}
\right.
\end{equation}
where $I_{\mathrm{N}}(t)$, $V_{\mathrm{N}}(t)$, $V_1(t)$, $V_2(t)$ are the zero-mean stochastic signals from corresponding noise sources $i_n$, $e_n$, $v_1$ and $v_2$.
We perform a small-signal analysis for the current fluctuations $\delta I$. \cref{eq:kirch_1} is then rewritten as
\begin{equation}
\left\{
\begin{aligned}
-e_n + v_1 &= R_{\mathrm{in}}\,\delta I + \delta V_{\mathrm{SD}},\\
\delta V_{\mathrm{out}} - v_2 - e_n &= R_{\mathrm{FB}}\,\delta I + R_{\mathrm{FB}}\, i_n \, .
\end{aligned}
\right.
\end{equation}
Rearranging, we obtain
\begin{subequations}\label{eq:3}
\begin{align}
\delta V_{\mathrm{SD}} &= -R_{\mathrm{in}}\,\delta I - e_n + v_1, \label{eq:3a}\\
\frac{\delta V_{\mathrm{out}}}{R_{\mathrm{FB}}} &= \frac{1}{R_{\mathrm{FB}}}\left(v_2 + e_n\right) + \delta I + i_n . \label{eq:3b}
\end{align}
\end{subequations}

\begin{figure}[t!]
  \centering
  \includegraphics[width=\textwidth]{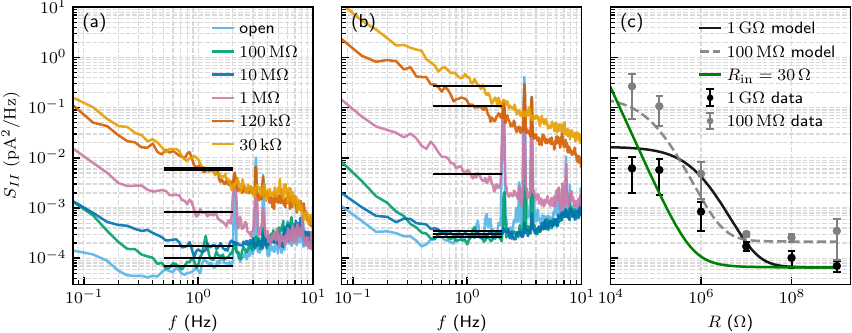}
  \caption{Characterization of the TIA using nominal resistive loads \(R\) cooled to \(\SI{300}{\milli\kelvin}\). Open (Coulomb blockade) condition is imitated by cooling down a highly resistive Si chip. (a) Current PSD \(S_{II}\) for the configuration \(R_{\mathrm{in}}=\SI{1}{\mega\ohm}\) and \(R_{\mathrm{FB}}=\SI{1}{\giga\ohm}\). (b) Current PSD \(S_{II}\) for the configuration \(R_{\mathrm{in}}=\SI{100}{\kilo\ohm}\) and \(R_{\mathrm{FB}}=\SI{100}{\mega\ohm}\). In both panels, the black horizontal lines mark the frequency range used to extract the average TIA noise level around \(\SI{1}{\hertz}\) with its value represented by the line's position on the $y$-axis. (c) Extracted noise at \(\SI{1}{\hertz}\) as a function of device resistance \(R\). Black and gray dots correspond to the \(R_{\mathrm{FB}}=\SI{1}{\giga\ohm}\) and \(R_{\mathrm{FB}}=\SI{100}{\mega\ohm}\) configurations, respectively. The solid black and gray lines show theoretical predictions based on \cref{eq:TIA_PSD}, with \(e_n = \SI{5}{\nano\volt}/\sqrt{\SI{}{\hertz}}\) and \(i_n = \SI{7}{\femto\ampere}/\sqrt{\SI{}{\hertz}}\). The green solid curve shows the corresponding prediction for a TIA with low input resistance $R_{\mathrm{in}} = \SI{30}{\ohm}$.}
  \label{fig:figure_TIA_noise}
\end{figure}

For current fluctuations, we have
\begin{equation}
\delta I\!\left(V_{\mathrm{SD}},\mu\right)
=
\left(\frac{\partial I}{\partial V_{\mathrm{SD}}}\right)_{\mu}\,\delta V_{\mathrm{SD}}
+
\left(\frac{\partial I}{\partial \mu}\right)_{V_{\mathrm{SD}}}\,\delta \mu \, .
\end{equation}
Plugging $\delta V_{\mathrm{SD}} = -R_{\mathrm{in}}\,\delta I - e_n + v_1$ from \cref{eq:3a} into the expression for $\delta I$, we obtain
\begin{equation}
\left(1 + R_{\mathrm{in}}\left(\frac{\partial I}{\partial V_{\mathrm{SD}}}\right)_{\mu}\right)\delta I
=
\left(\frac{\partial I}{\partial V_{\mathrm{SD}}}\right)_{\mu}\left(v_1 - e_n\right)
+
\left(\frac{\partial I}{\partial \mu}\right)_{V_{\mathrm{SD}}}\,\delta \mu \, .
\end{equation}
Note that $\left(\dfrac{\partial I}{\partial V_{\mathrm{SD}}}\right)_{\mu}=G$ is the conductance of the QD.
Promoting this expression to the power spectral density, we obtain
\begin{equation}
S_{II}
=
\left(1+R_{\mathrm{in}}G\right)^{-2}
\left[
G^{2}\left(v_{1}^{2}+e_{\mathrm{n}}^{2}\right)
+
\left(\frac{\partial I}{\partial \mu}\right)_{V_{\mathrm{SD}}}^{2} \, S_{\mu\mu}
\right]. \label{eq:curr_PSD}
\end{equation}
The power spectral density of the total measured signal $\delta V_{\mathrm{out}}/R_{\mathrm{FB}}$ (from \cref{eq:3b}) is
\begin{equation}
S_{II}^{\mathrm{tot}}
= i_{\mathrm{n}}^{2}
+
\frac{1}{R_{\mathrm{FB}}^{2}}\left(v_2^2+e_{\mathrm{n}}^{2}\right) + S_{II}.\label{eq:tot_PSD}
\end{equation}
Plugging \cref{eq:curr_PSD} into \cref{eq:tot_PSD} and rearranging the terms, we obtain
\begin{equation}
S_{II}^{\mathrm{tot}}
=
S_{II}^{\mathrm{TIA}}
+
S_{II}^{\mathrm{device}}, \label{eq:final_PSD}
\end{equation}
where
\begin{equation}
S_{II}^{\mathrm{TIA}}
=
\underbrace{
i_{\mathrm{n}}^{2}
+
\frac{1}{R_{\mathrm{FB}}^{2}}\left(4k_{\mathrm{B}}TR_{\mathrm{FB}}+e_{\mathrm{n}}^{2}\right)
}_{S\ \text{in Coulomb blockade}}
+
\left(\frac{G}{1+R_{\mathrm{in}}G}\right)^{2}
\left(e_{\mathrm{n}}^{2}+4k_{\mathrm{B}}TR_{\mathrm{in}}\right),\label{eq:TIA_PSD}
\end{equation}
and
\begin{equation}
S_{II}^{\mathrm{device}}
=
\left(1+R_{\mathrm{in}}G\right)^{-2}
\left(\frac{\partial I}{\partial \mu}\right)_{V_{\mathrm{SD}}}^{2}
S_{\mu\mu}.\label{eq:device_PSD}
\end{equation}

From \cref{eq:TIA_PSD}, we immediately see that the TIA noise level depends on the differential conductance of the QD. In the Coulomb blockade regime, where \(G=0\), the third term in \cref{eq:TIA_PSD} vanishes, whereas at high conductance its contribution is limited by \(R_{\mathrm{in}}\). To test this prediction experimentally, we cooled to \(\SI{300}{\milli\kelvin}\) a set of resistors with nominal values \(R = \SI{30}{\kilo\ohm},\ \SI{120}{\kilo\ohm},\ \SI{1}{\mega\ohm},\ \SI{10}{\mega\ohm},\ \SI{100}{\mega\ohm}\). To emulate the Coulomb blockade limit, we additionally bonded a highly resistive Si chip, which becomes insulating at low temperature. Using the same PSD estimation procedure as for the QD measurements, we acquired current-noise spectra. The results are shown in \cref{fig:figure_TIA_noise}a and b for the configurations \(R_{\mathrm{in}}=\SI{1}{\mega\ohm}\), \(R_{\mathrm{FB}}=\SI{1}{\giga\ohm}\) and \(R_{\mathrm{in}}=\SI{100}{\kilo\ohm}\), \(R_{\mathrm{FB}}=\SI{100}{\mega\ohm}\), respectively. The TIA-added noise at \(\SI{1}{\hertz}\) is estimated by averaging the spectral density over the frequency range marked by the solid black lines. These values are plotted as a function of device resistance \(R\) in \cref{fig:figure_TIA_noise}c.

The experimental data are in qualitative agreement with the theoretical curves (black and gray) calculated from \cref{eq:TIA_PSD}, where we set \(G = 1/R\). The exact value of the voltage-noise source \(e_n\) has only a weak effect on the fit, and we therefore fix it to a value \(e_n = \SI{5}{\nano\volt}/\sqrt{\SI{}{\hertz}}\), representative for TIA. The current-noise source is set to \(i_n = \SI{7}{\femto\ampere}/\sqrt{\SI{}{\hertz}}\), consistent with the value specified by the TIA manufacturer. Importantly, 713 of the 761 traces considered in this work have QD noise values at \(\SI{1}{\hertz}\) that exceed the maximal saturated TIA-added noise. For the remaining traces, separating the device contribution from the amplifier background would require a more careful conductance-based analysis. We do not perform such an analysis here, since these traces do not affect the overall statistics and do not lead to an underestimation of the reported noise values. For comparison, we also plot the theoretical curve for a TIA with the same noise parameters but a low input resistance, \(R_{\mathrm{in}} = \SI{30}{\ohm}\)~\href{https://www.baspi.ch/low-noise-high-stab-itov-conv}{(\eg \textit{https://www.baspi.ch/low-noise-high-stab-itov-conv)}}.

We now turn to the device noise contributions, described by \cref{eq:TIA_PSD}. The prefactors in front of $S_{\mu\mu}$ differ from those appearing in the literature\cite{connorsLowfrequencyChargeNoise2019a}. This is the difference between performing the experiment at fixed external $V_{\mathrm{bias}}$ voltage and fixed voltage drop over the QD $V_{\mathrm{SD}}$. We show how to convert from one case to another. For the prefactors we can work with the mean values, and therefore the noise sources will average out. For the mean values, we have 

\begin{equation}
V_{\mathrm{bias}} = V_{\mathrm{SD}} + R_{\mathrm{in}}\, I\!\left(V_{\mathrm{SD}},\mu\right) = \mathrm{const}.
\label{eq:Vbias_def}
\end{equation}
Eq.~\eqref{eq:Vbias_def} defines an implicit path $V_{\mathrm{SD}}=V_{\mathrm{SD}}(\mu)$ at fixed $V_{\mathrm{bias}}$.
Along this path,
\begin{align}
\left(\frac{\partial I}{\partial \mu}\right)_{V_{\mathrm{bias}}}
&=
\frac{d}{d\mu} I\!\bigl(V_{\mathrm{SD}}(\mu),\mu\bigr) \nonumber\\
&=
\left(\frac{\partial I}{\partial V_{\mathrm{SD}}}\right)_{\mu}\frac{dV_{\mathrm{SD}}}{d\mu}
+
\left(\frac{\partial I}{\partial \mu}\right)_{V_{\mathrm{SD}}} .
\label{eq:chain_rule}
\end{align}
Differentiate Eq.~\eqref{eq:Vbias_def}:
\begin{equation}
dV_{\mathrm{bias}}
=
\Bigl(1+R_{\mathrm{in}}\left(\frac{\partial I}{\partial V_{\mathrm{SD}}}\right)_{\mu}\Bigr)\,dV_{\mathrm{SD}}
+
R_{\mathrm{in}}\left(\frac{\partial I}{\partial \mu}\right)_{V_{\mathrm{SD}}}\,d\mu .
\label{eq:dVbias}
\end{equation}
For fixed $V_{\mathrm{bias}}$ ($\frac{dV_{\mathrm{bias}}}{d\mu}=0$),
\begin{equation}
\frac{dV_{\mathrm{SD}}}{d\mu}
=
-\,R_{\mathrm{in}}
\left(\frac{\partial I}{\partial \mu}\right)_{V_{\mathrm{SD}}}
\Bigl(1+R_{\mathrm{in}}\left(\frac{\partial I}{\partial V_{\mathrm{SD}}}\right)_{\mu}\Bigr)^{-1}.
\label{eq:dVsd_dmu}
\end{equation}
Insert Eq.~\eqref{eq:dVsd_dmu} into Eq.~\eqref{eq:chain_rule}:
\begin{equation}
\left(\frac{\partial I}{\partial \mu}\right)_{V_{\mathrm{bias}}}
=
\left(\frac{\partial I}{\partial \mu}\right)_{V_{\mathrm{SD}}}
\Bigl(1+R_{\mathrm{in}}\left(\frac{\partial I}{\partial V_{\mathrm{SD}}}\right)_{\mu}\Bigr)^{-1}.
\label{eq:dIdmu_relation}
\end{equation}

Plugging Eq.~\eqref{eq:dIdmu_relation} into \cref{eq:device_PSD}, and assigning the electrochemical potential to the gate voltage $\mu = \alpha V_{\mathrm{g}}$ ($\alpha$ is the lever arm), we obtain:
\begin{equation}
S_{II}^{\mathrm{device}}
=
\alpha^{-2}\left(\frac{\partial I}{\partial V_{\mathrm{g}}}\right)_{V_{\mathrm{bias}}}^{\!2}
S_{\mu\mu},
\end{equation}
or, finally, as used in the main text:
\begin{equation}
S_{\mu\mu}
=
\alpha^{2}\left(\frac{\partial I}{\partial V_{\mathrm{g}}}\right)_{V_{\mathrm{bias}}}^{\!-2}
S_{II}.
\end{equation}

\newpage
\section{Noise simulations}\label{sec:simulations}
The purpose of the present section is to assess the limitations of the method used in the core of this work. We do not cover microscopic simulations of TLFs (for this, we refer interested readers to Ref.\onlinecite{shehataModelingSemiconductorSpin2023}); instead, we demonstrate how the method works on synthetic noise datasets generated using toy-models.

We proceed as follows. Without loss of generality, we assume that the current $I(V_{\mathrm{g}})$ over a Coulomb peak is limited by the life-time broadening and is given by the expression\cite{kouwenhovenElectronTransportQuantum1997}:
\begin{equation}
    I(V_{\mathrm{g}}) = \frac{I_0}{1 + \left(\frac{\alpha V_{\mathrm{g}}}{h\Gamma}\right)^2},
\end{equation}
$h$ is the Planck constant. To make the simulated dataset look like the data in Fig.2 of the main text, we choose $\alpha = \SI{0.085}{\electronvolt}/\SI{}{\volt}$, $\Gamma \sim \SI{25}{\giga \hertz} \gtrsim \SI{20}{\giga \hertz} = \SI{100}{\milli \kelvin}$, and $I_0 \sim \SI{90}{\pico \ampere}$. The noise is simulated by generating time traces with a given PSD model (Lorentz or $1/f$) for voltage fluctuations $V_{\mathrm{g}}(t) = V_{g} + \delta V_{\mathrm{g}}(t)$. The case of the sum model can be analyzed similarly. We add the white-noise component to the current to mimic the background noise $I(t) = I(V_{g}(t)) \approx I(V_{\mathrm{g}}) + \left.\frac{dI}{dV}\right|_{V_{\mathrm{g}}}\delta V(t) + \delta I(t)$. As in the experiment, we set the time $T = \SI{40}{\second}$ and the number of samples $N = 50000$.

\subsection{Single TLF (Lorentz model).}
First, we simulate the case of a single TLF. We consider voltage fluctuations as switches between two levels $-\delta V_{\mathrm{g}}/2$ and $\delta V_{\mathrm{g}}/2$ with the rates $1/\tau_{in}$ and $1/\tau_{out}$. The telegraphic nature of the fluctuations is usually modeled by a two-state continuous time Markov chain. The Markov ('memoryless') property is equivalent to saying that the dwelling times between switches are exponentially distributed, \ie are drawn from the distribution with the probability density function $\frac{1}{\tau}e^{-t/\tau}$ (see Ref. \onlinecite{norrisMarkovChains1997}). In this case, to simulate the random telegraphic noise, it suffices to randomly choose from the exponential distribution switching times $T_1$, $T_2$, ..., until the end of the time trace $T_1 + T_2 + ... + T_n \leq T$. The PSD of such a Markov process is then given by the Lorentz formula\cite{koganElectronicNoiseFluctuations1996}:
\begin{equation}
    S_{VV}(f) = 4\frac{\tau_{in}\tau_{out}}{(\tau_{in} + \tau_{out})^2}\frac{\tau_c(\delta V_{\mathrm{g}})^2}{1 + (2\pi f\tau_c)^2},
    \label{eq:lorentz_theor}
\end{equation}
where $\tau_c^{-1} = \tau_{in}^{-1} + \tau_{out}^{-1}$. 

\begin{figure}[t!]
  \centering
  \includegraphics[width=\textwidth]{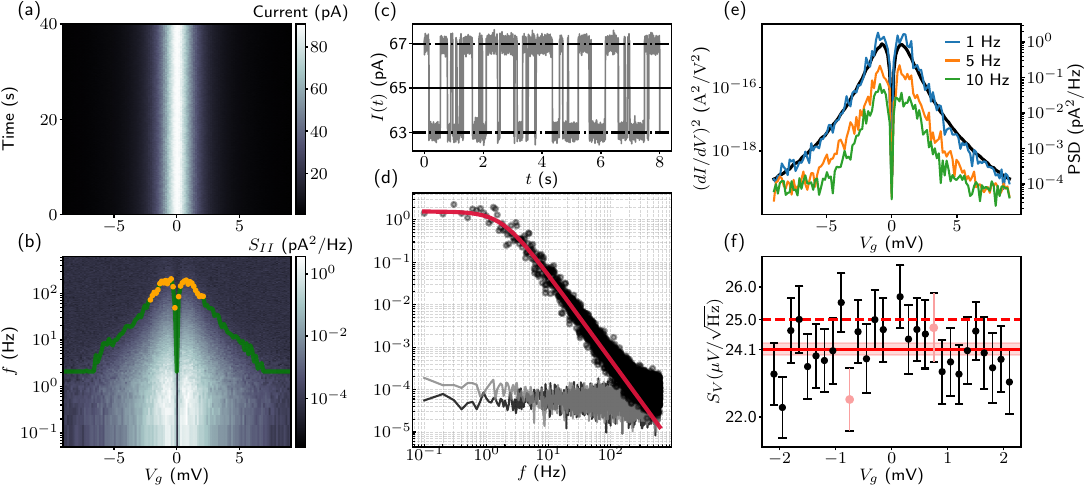}
  \caption{Simulated single TLF noise. (a) time traces over the Coulomb peak. (b) The PSD estimated with Bartlett method $N_{\textrm{seg}} = 4$. The overlapped green curve corresponds to frequencies $f^*$ where the device and background spectra become indistinguishable. The orange highlight the selected for the analysis points. (c) Current trace $I(t)$ taken at the $V_{\mathrm{g}}$ correspondent to maximal derivative. Solid line is set at $I(V_{\mathrm{g}})$ value, while the dotted lines are precisely $I = I(V_{\mathrm{g}}) \pm \frac{dI}{dV}\delta V / 2$. (d) Example PSD spectra: dots - at maximal derivative, grey - maximal current, black - minimal current. The red trace shows the Lorentz fit. (e) Cuts in the PSD 2D plot at 1Hz, 5Hz, 10Hz. The black trace is the square of the numerical derivative of the averaged current (in log-scale). (f) Black dots with error bars are extracted from the fits noise values at 1Hz. Red line with error bar is the averaged value. Dashed red line is the value plugged into the simulation. The pink points highlight the values at voltages with highest sensitivity.}
  \label{fig:figureS6}
\end{figure}

The form of \cref{eq:lorentz_theor} is exactly $\frac{B}{(f/f_0)^2 + 1}$, which we use for fitting. For our simulation we set $\delta V_{\mathrm{g}}$, such that the resulting voltage noise at $S_V(\SI{1}{\hertz})$, calculated from \cref{eq:lorentz_theor}, is approximately the one we observed in Fig. 2 of the main text. For $f_0 = 1/(2\pi \tau_c)$, we choose the values $\tau_{in} = \tau_{out} = \SI{0.3}{\second}$, to roughly match the experimental results.

In \cref{fig:figureS6}a we plot the time traces with the synthetic noise and in \cref{fig:figureS6}b the PSD spectra over the Coulomb peak. The single TLF noise nature can be observed from \cref{fig:figureS6}c, where the time trace is taken as an example at the $V_{\mathrm{g}}$ of maximum sensitivity. The solid line highlights $I(V_{\mathrm{g}})$ and two dotted lines are precisely $I = I(V_{\mathrm{g}}) \pm \frac{dI}{dV}\delta V / 2$. In \cref{fig:figureS6}b we observe enhancement of the noise levels where the sensitivity is increased. This can be seen very well in \cref{fig:figureS6}e, where the noise cuts at certain frequencies directly follow the squared sensitivity $(dI/dV_{\mathrm{g}})^2$ (log-scale).

In this dataset, we choose those traces for the fit that are distinguishable from the background, \ie noise spectrum at the apex of the Coulomb peak, at frequencies below 50Hz (for real devices the cutoff is chosen smaller due to the roll-off of the transimpedance apmlifier). The frequencies for each gate voltage $f^*(V_{\mathrm{g}})$, at which the device-noise PSD becomes indistinguishable from the background are plotted as the green line in \cref{fig:figureS6}. An example of the distinguishable trace is given in \cref{fig:figureS6}d. The selected points are marked with orange in \cref{fig:figureS6} and the corresponding PSDs are following Lorentz trends. From the fits we extract current PSD $S_{II}$. We convert $S_{II}$ to the voltage PSD by using formula $S_{VV} = \left(\frac{dI}{dV}\right)^{-2} S_{II}$. We plot extracted values with their error-bars from the fitting in \cref{fig:figureS6}f. We see that the average value (red solid line) together with weighted error-bars is below the value we put into simulation (red dotted line). We will come to this point later after showing similar analysis for the distribution of TLFs. We would like to note here two features. First, there is no trend in how the extracted noise datapoints are distributed across the Coulomb peak. Second, the points with maximal sensitivities (pink) are not yielding more precise values than the rest.

\subsection{Distribution of TLFs ($1/f$ model).}
Next we proceed with the simulation of $1/f$ spectra. Rather than sampling single TLFs within a certain frequency band, we decided instead to convert the $1/f$ spectra into the time trace $V_{\mathrm{g}}(t)$. The algorithm on how to do it is described in detail in Ref.\onlinecite{timmerGeneratingPowerLaw1995}. First, the Fourier components of the random signal are drawn from the normal distribution, \ie $c(f) = a(f) + i b(f)$, such that $a(f)$, $b(f) \sim N(0,1)$. Then, the signal is convoluted with the desired PSD $\int S(f)c(f-\nu)d\nu$, where $S(f) = A/f^{\beta}$, and the inverse Fourier transform is applied.

For the single TLF model we were synthesizing fluctuations for each $V_{\mathrm{g}}$ independently. This is justified, because for the frequencies smaller $f_0$ the noise power is saturated, \ie the system does not have low-frequency drifts. For $1/f$ more care should be taken to include these drifts. As such, we generate the noise time trace for the duration of the entire experiment. In other words we synthesize the $1/f$ noise up to the $1 / T_{\textrm{full}}$ frequency, and then take segments of duration $T$ for each of the plunger voltages $V_{\mathrm{g}}$. Like this we would also take into account low-frequency drifts of the Coulomb peak during the experiment.

\begin{figure}[t!]
  \centering
  \includegraphics[width=\textwidth]{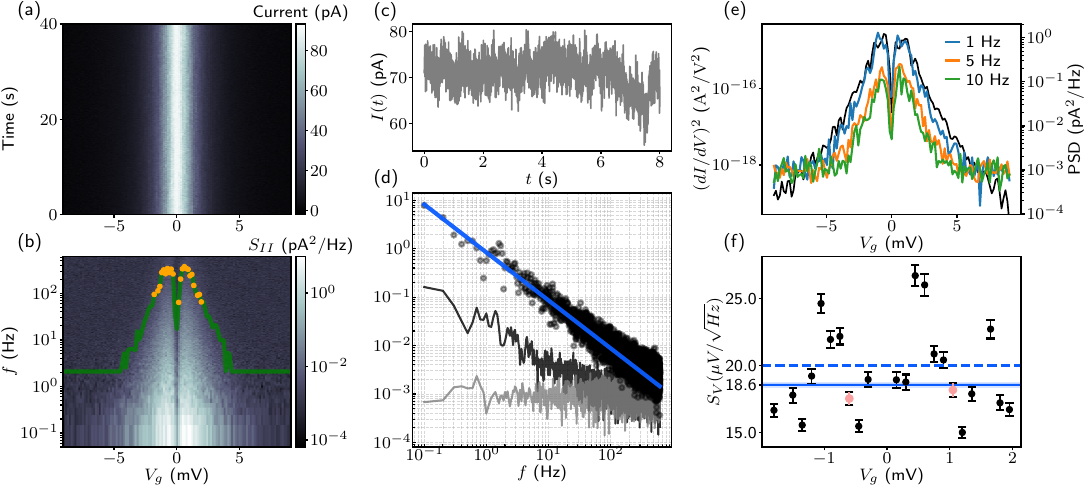}
  \caption{Simulated $1/f$ noise. Here the amplitude is set to $A = \SI{20}{\micro\volt}$. (a) time traces over the Coulomb peak. (b) The PSD estimated with Bartlett method $N_{\textrm{seg}} = 4$. The overlapped green curve corresponds to frequencies $f^*$ where the device and background spectra become indistinguishable. The orange highlight the selected for the analysis points. (c) Current trace $I(t)$ taken at the $V_{\mathrm{g}}$ correspondent to the maximal derivative. (d) Example PSD spectra: dots - at maximal derivative, grey - maximal current, black - minimal current. The blue trace shows the $1/f$ fit. (e) Cuts in the PSD 2D plot at 1Hz, 5Hz, 10Hz. The black trace is the square of the numerical derivative of the averaged current (in log-scale). (f) Black dots with error bars are extracted from the fits noise values at 1Hz. The blue line with the error bar is the averaged value. The dashed blue line is the value plugged into the simulation. The pink points highlight the values at voltages with highest sensitivity.}
  \label{fig:figureS7}
\end{figure}

For demonstration purposes we choose $\beta=1$ and $A = \SI{20}{\micro\volt}$. It does not match the experimental data in Fig.2 of the main text, however, showcases the situation when noise is larger $S_{\mu} \approx \SI{2}{\micro\electronvolt}/\sqrt{\SI{}{\hertz}}$ ($\alpha \sim 0.1 \si{\electronvolt}/\si{\volt}$). The analysis is analogous to the one before and is presented in \cref{fig:figureS7}. We see that the generated noise (\cref{fig:figureS7}c) is exactly $1/f$-trend (\cref{fig:figureS7}d) with $\beta = 1$ and the noise cuts (\cref{fig:figureS7}e) follow the derivative. We note that the black trace, taken at the apex of the Coulomb peak, also exhibits an approximately $1/f$ trend. We attribute this to slow gate-voltage drift: low-frequency fluctuations shift the effective operating point away from the peak maximum, so the nominal “apex” condition samples finite-slope regions of the peak.
\begin{figure}[t!]
  \centering
  \includegraphics[width=\textwidth]{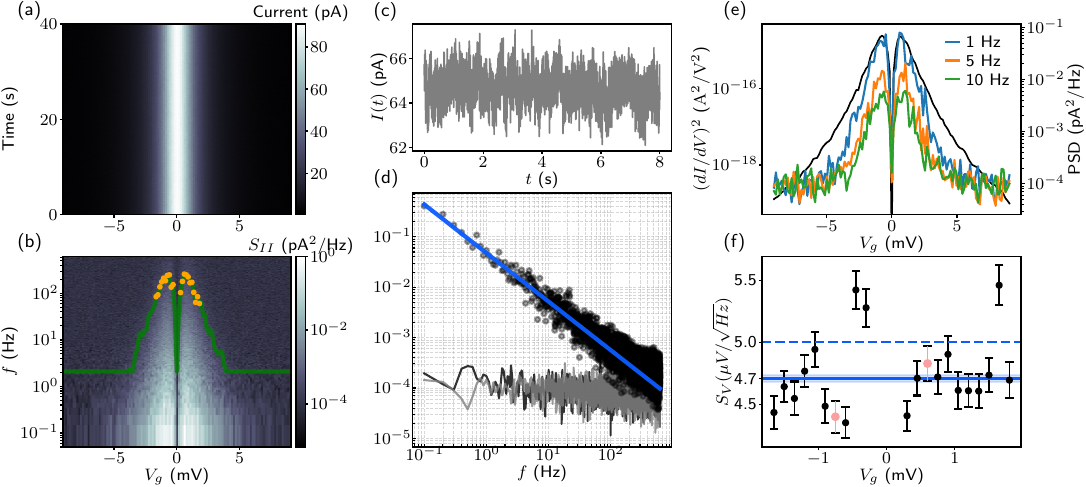}
  \caption{Simulated $1/f$ noise. Here the amplitude is set to $A = \SI{5}{\micro\volt}$. (a) time traces over the Coulomb peak. (b) The PSD estimated with Bartlett method $N_{\textrm{seg}} = 4$. The overlapped green curve corresponds to frequencies $f^*$ where the device and background spectra become indistinguishable. The orange highlight the selected for the analysis points. (c) Current trace $I(t)$ taken at the $V_{\mathrm{g}}$ correspondent to the maximal derivative. (d) Example PSD spectra: dots - at maximal derivative, grey - maximal current, black - minimal current. The blue trace shows the $1/f$ fit. (e) Cuts in the PSD 2D plot at 1Hz, 5Hz, 10Hz. The black trace is the square of the numerical derivative of the averaged current (in log-scale). (f) Black dots with error bars are extracted from the fits noise values at 1Hz. The blue line with the error bar is the averaged value. The dashed blue line is the value plugged into the simulation. The pink points highlight the values at voltages with highest sensitivity.}
  \label{fig:figureS8}
\end{figure}
Another observation is important: the squared sensitivity (\cref{fig:figureS7}e) exhibits pronounced fluctuations. This directly contributes to the spread of the extracted noise values at $\SI{1}{\hertz}$, even though this spread is not limited by the PSD estimation uncertainty (the error bars are small compared to the overall dispersion). This observation highlights an intrinsic limitation of flank-method-based noise estimation, which we formulate as following. Similar conclusions were derived in Ref.\cite{holmanCircuitQuantumElectrodynamics}.

The uncertainty of the PSD estimation with the periodogram approach can be reduced by increasing the number of averaged segments $N_{\mathrm{seg}}$. This, however, can only be achieved at the expense of either increasing the total acquisition time $T$ or decreasing the segment length $T_{\mathrm{seg}}$, which increases the minimal frequency bin $f_{\mathrm{min}} \sim 1 / T_{\mathrm{seg}}$. Increasing $T$ in turn makes the experiment more susceptible to low-frequency drifts: slow fluctuations shift the effective operating point along the Coulomb peak and hence change the local sensitivity to higher-frequency fluctuations. Therefore, only limited precision is achievable for reasonable charge noise and duration of the experiment. 

To finalize, we present the similar simulation for the charge noise value comparable to the one in Fig.2 of the main text. We see in \cref{fig:figureS8}e that the spread of the estimated values dropped compared to the larger noise value put into the simulation.

\clearpage
\section{Summary of all the datasets}\label{app:summary}
\begin{figure}[t!]
  \centering
  \includegraphics[width=\textwidth]{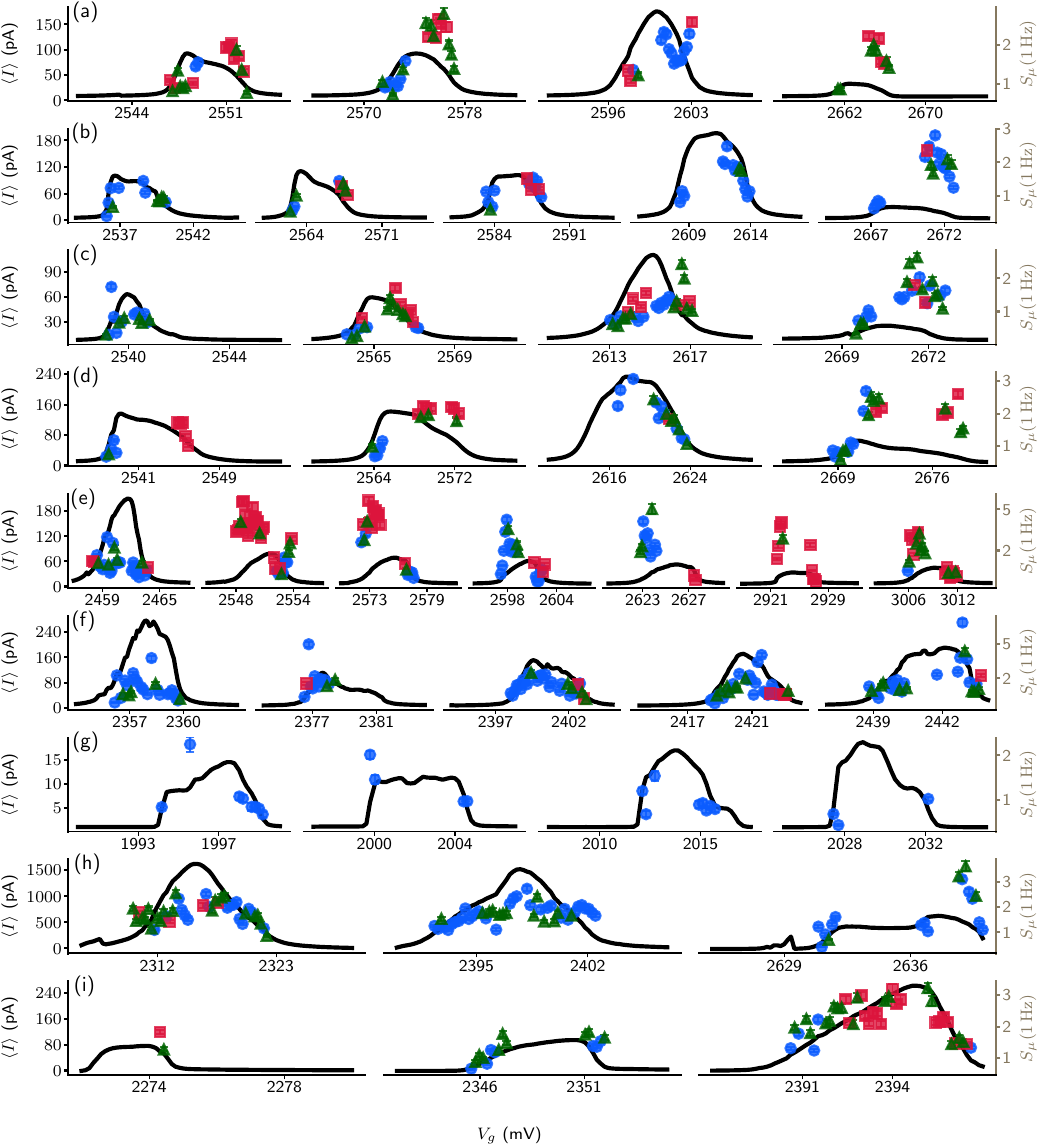}
  \caption{\footnotesize{Summary of analyzed Coulomb peaks in this study under different TIA and bias settings. (a) Device A, Left Dot, $V_{\mathrm{bias}}=\SI{270}{\micro\volt}$ and $R_{\mathrm{FB}}=\SI{1}{\giga\ohm}$. (b) Same device and dot with $R_{\mathrm{FB}}=\SI{100}{\mega\ohm}$ (other conditions nominally unchanged). (c) Device A, Left Dot, $R_{\mathrm{FB}}=\SI{1}{\giga\ohm}$ and $V_{\mathrm{bias}}=\SI{40}{\micro\volt}$. (d) Same device and dot with $V_{\mathrm{bias}}=\SI{640}{\micro\volt}$. (e) Device A, Right Dot. (f) Device A, Sensor Dot. (g) Device B. (h) Device C, measured with $R_{\mathrm{FB}}=\SI{100}{\mega\ohm}$. (i) Same device with $R_{\mathrm{FB}}=\SI{1}{\giga\ohm}$.}}
  \label{fig:SI_summary_coulomb}
\end{figure}

In \cref{fig:SI_summary_coulomb} we collected all datasets analyzed in the present study. For visual guidance, we plot the average current $\left<I\right>$ across a Coulomb peak (the current values are on the left axis) together with the extracted charge noise at $\SI{1}{\hertz}$ $S_{\mu}$ (the noise values on the right axis). 

Rows (a)–(d) show measurements on the left QD of Device A acquired under different TIA configurations and biases. Between rows (a) and (b), we changed the TIA feedback resistor $R_{\mathrm{FB}}$ from $\SI{1}{\giga \ohm}$ to $\SI{100}{\mega \ohm}$. For the TIA we were using during the experiments, change in the feedback resistance $R_{\mathrm{FB}}$ -- on the hardware level, see ~\href{https://qtwork.tudelft.nl/~schouten/ivvi/doc-mod/docm1b.htm}{\textit{https://qtwork.tudelft.nl/~schouten/ivvi/doc-mod/docm1b.htm}} -- implies change in the input resistance $R_{\mathrm{in}}$: it changes from $\SI{1}{\mega \ohm}$ to $\SI{100}{\kilo \ohm}$. Because the electrostatic operating point shifted between the two measurements, the comparison is not strictly one-to-one; nevertheless, qualitatively the noise levels remain unchanged. An analogous test on Device C (rows (h) and (i)) leads to the same conclusion, indicating that the spread of the extracted noise is likely coming from the local electrostatic conditions. 

We performed all subsequent measurements with a $\SI{1}{\giga \ohm}$ feedback resistor. In rows (c) and (d), we vary the bias voltage from $V_{\mathrm{bias}} = \SI{40}{\micro\volt}$ to $V_{\mathrm{bias}} = \SI{640}{\micro\volt}$, again without a systematic change in the averaged noise values. Finally, rows (e)–(g) correspond to the right dot of Device A, the sensor dot of Device A, and Device B, respectively. The bias was set around $V_{\mathrm{bias}} = \SI{400}{\micro\volt}$.

\end{document}